\documentclass[a4paper]{article}
\usepackage{amssymb}
\usepackage{amsfonts}
\usepackage{amsmath}
\usepackage[margin=2cm]{geometry}
\usepackage{graphicx}
\usepackage{caption}
\usepackage{subfig}
\usepackage{multirow}
\usepackage{multicol}

\input{tcilatex}

\begin{document}

\title{Fractional Burgers wave equation}
\author{ Ljubica Oparnica\thanks{
Faculty of Education, University of Novi Sad, Podgori\v cka 4, 25000 Sombor,
Serbia and Department of Mathematics: Analysis, Logic and Discrete
Mathematics, University of Gent, Krijgslaan 281 (building S8), 9000 Gent,
Belgium, Oparnica.Ljubica@UGent.be}, Du\v{s}an Zorica\thanks{
Mathematical Institute, Serbian Academy of Arts and Sciences, Kneza Mihaila
36, 11000 Belgrade, Serbia and Department of Physics, Faculty of Sciences,
University of Novi Sad, Trg D. Obradovi\'{c}a 4, 21000 Novi Sad, Serbia,
dusan\textunderscore zorica@mi.sanu.ac.rs}, Aleksandar S. Okuka\thanks{
Department of Mechanics, Faculty of Technical Sciences, University of Novi
Sad, Trg D. Obradovi\'{c}a 6, 21000 Novi Sad, Serbia, aokuka@uns.ac.rs}}
\maketitle

\begin{abstract}
\noindent Thermodynamically consistent fractional Burgers constitutive
models for viscoelastic media, divided into two classes according to model
behavior in stress relaxation and creep tests near the initial time instant,
are coupled with the equation of motion and strain forming the fractional
Burgers wave equations. Cauchy problem is solved for both classes of Burgers
models using integral transform method and analytical solution is obtained
as a convolution of the solution kernels and initial data. The form of
solution kernel is found to be dependent on model parameters, while its
support properties implied infinite wave propagation speed for the first
class and finite for the second class. Spatial profiles corresponding to the
initial Dirac delta displacement with zero initial velocity display features
which are not expected in wave propagation behavior.

\end{abstract}

\section{Introduction}

Fractional Burgers wave equation is written as the system of equations
consisting of: equation of motion corresponding to one-dimensional
deformable body%
\begin{equation}
\frac{\partial }{\partial x}\sigma (x,t)=\rho \,\frac{\partial ^{2}}{%
\partial t^{2}}u(x,t),\;\;x\in \mathbb{R},\;t>0,  \label{eq-motion}
\end{equation}%
where $u$ and $\sigma $ are displacement and stress, while $\rho $ is
constant material density; strain for small local deformations%
\begin{equation}
\varepsilon (x,t)=\frac{\partial }{\partial x}u(x,t),\;\;x\in \mathbb{R}%
,\;t>0;  \label{strejn}
\end{equation}%
and constitutive equation represented by the fractional Burgers model%
\begin{equation}
\left( 1+a_{1}\,{}_{0}\mathrm{D}_{t}^{\alpha }+a_{2}\,{}_{0}\mathrm{D}%
_{t}^{\beta }+a_{3}\,{}_{0}\mathrm{D}_{t}^{\gamma }\right) \sigma \left(
x,t\right) =\left( b_{1}\,{}_{0}\mathrm{D}_{t}^{\mu }+b_{2}\,{}_{0}\mathrm{D}%
_{t}^{\nu }\right) \varepsilon \left( x,t\right) ,\;\;x\in \mathbb{R},\;t>0,
\label{fbm}
\end{equation}%
having model parameters assumed as: $a_{1},a_{2},a_{3},b_{1},b_{2}>0,$ $%
\alpha ,\beta ,\mu \in \left[ 0,1\right] ,$ with $\alpha \leq \beta ,$ and $%
\gamma ,\nu \in \left[ 1,2\right] ,$ while the operator of Riemann-Liouville
fractional derivative ${}_{0}\mathrm{D}_{t}^{\xi }$ of order $\xi \in \left[
n,n+1\right] ,$ $n\in 
\mathbb{N}
_{0},$ is defined by%
\begin{equation*}
{}_{0}\mathrm{D}_{t}^{\xi }y\left( t\right) =\frac{\mathrm{d}^{n+1}}{\mathrm{%
d}t^{n+1}}\left( \frac{t^{-\left( \xi -n\right) }}{\Gamma \left( 1-\left(
\xi -n\right) \right) }\ast y\left( t\right) \right) ,\;\;t>0,
\end{equation*}%
see \cite{TAFDE}, where $\ast $ denotes the convolution in time: $f\left(
t\right) \ast _{t}g\left( t\right) =\int_{0}^{t}f\left( t^{\prime }\right)
g\left( t-t^{\prime }\right) \mathrm{d}t^{\prime },$ $t>0.$

In order to solve the Cauchy problem on the real line $x\in \mathbb{R}$ and $%
t>0,$ the system of governing equations (\ref{eq-motion}), (\ref{strejn}),
and (\ref{fbm}) is subject to initial and boundary conditions: 
\begin{gather}
u(x,0)=u_{0}(x),\;\;\frac{\partial }{\partial t}u(x,0)=v_{0}(x),\;\;\sigma
(x,0)=0,\;\;\varepsilon (x,0)=0,  \label{ic} \\
\lim_{x\rightarrow \pm \infty }u(x,t)=0,\;\;\lim_{x\rightarrow \pm \infty
}\sigma (x,t)=0,  \label{bc}
\end{gather}%
where $u_{0}$ is the initial displacement and $v_{0}$ is the initial
velocity.

Considering the rheological scheme of the classical Burgers model, with the
dash-pot element replaced by the Scott-Blair (fractional) element, the
fractional Burgers model (\ref{fbm}) is derived in \cite{OZ-1}. Moreover,
using the requirement of storage and loss modulus non-negativity, the
analysis of thermodynamical consistency for fractional Burgers model (\ref%
{fbm}), conducted in \cite{OZ-1}, yielded that the orders of fractional
derivatives $\gamma ,\nu \in \left[ 1,2\right] $ cannot be independent of
the orders of fractional derivatives $\alpha ,\beta ,\mu \in \left[ 0,1%
\right] ,$ and this led to formulation of eight thermodynamically consistent
fractional Burgers models, divided into two classes.

The first class contains five models, written as 
\begin{equation}
\left( 1+a_{1}\,{}_{0}\mathrm{D}_{t}^{\alpha }+a_{2}\,{}_{0}\mathrm{D}%
_{t}^{\beta }+a_{3}\,{}_{0}\mathrm{D}_{t}^{\gamma }\right) \sigma \left(
t\right) =\left( b_{1}\,{}_{0}\mathrm{D}_{t}^{\mu }+b_{2}\,{}_{0}\mathrm{D}%
_{t}^{\mu +\eta }\right) \varepsilon \left( t\right)  \label{UCE-1-5}
\end{equation}%
in an unified manner, such that the highest fractional differentiation order
of strain is $\mu +\eta \in \left[ 1,2\right] ,$ with $\eta \in \left\{
\alpha ,\beta \right\} ,$ while the highest fractional differentiation order
of stress is either $\gamma \in \left[ 0,1\right] $ in the case of Model I,
with $0\leq \alpha \leq \beta \leq \gamma \leq \mu \leq 1$ and $\eta \in
\left\{ \alpha ,\beta ,\gamma \right\} ,$ or $\gamma \in \left[ 1,2\right] $
in the case of Models II - V, with $0\leq \alpha \leq \beta \leq \mu \leq 1$
and $\left( \eta ,\gamma \right) \in \left\{ \left( \alpha ,2\alpha \right)
,\left( \alpha ,\alpha +\beta \right) ,\left( \beta ,\alpha +\beta \right)
,\left( \beta ,2\beta \right) \right\} $. The fractional differentiation
order of stress is less than the differentiation order of strain regardless
on the interval $\left[ 0,1\right] $ or $\left[ 1,2\right] .$

The second class contains three models, written as%
\begin{equation}
\left( 1+a_{1}\,{}_{0}\mathrm{D}_{t}^{\alpha }+a_{2}\,{}_{0}\mathrm{D}%
_{t}^{\beta }+a_{3}\,{}_{0}\mathrm{D}_{t}^{\beta +\eta }\right) \sigma
\left( t\right) =\left( b_{1}\,{}_{0}\mathrm{D}_{t}^{\beta }+b_{2}\,{}_{0}%
\mathrm{D}_{t}^{\beta +\eta }\right) \varepsilon \left( t\right)
\label{UCE-6-8}
\end{equation}%
in an unified manner, such that $0\leq \alpha \leq \beta \leq 1$ and $\beta
+\eta \in \left[ 1,2\right] ,$ with $\eta =\alpha ,$ in the case of Model
VI; $\eta =\beta $ in the case of Model VII; and $\alpha =\eta =\beta ,$ $%
\bar{a}_{1}=a_{1}+a_{2},$ and $\bar{a}_{2}=a_{3}$ in the case of Model VIII.
Considering the interval $\left[ 0,1\right] ,$ the highest fractional
differentiation orders of stress and strain are equal, which also holds true
for the orders from interval $\left[ 1,2\right] .$

The responses in creep and stress relaxation tests for Models I - VIII are
examined in \cite{OZ-2}. Recall, creep compliance $\varepsilon _{cr}$
(relaxation modulus $\sigma _{sr}$) is the strain (stress) history function
obtained as a response to the stress (strain) assumed as the Heaviside step
function. It is found that models' behavior near the initial time-instant is
different for the first and the second model class: Models I - V have zero
glass compliance, i.e., $\varepsilon _{cr}^{\left( g\right) }=\varepsilon
_{cr}\left( 0\right) =0$ and thus infinite glass modulus, i.e., $\sigma
_{sr}^{\left( g\right) }=\sigma _{sr}\left( 0\right) =\infty ,$ while Models
VI - VIII have non-zero glass compliance $\varepsilon _{cr}^{\left( g\right)
}=\frac{a_{3}}{b_{2}}$ implying the non-zero glass modulus $\sigma
_{sr}^{\left( g\right) }=\frac{b_{2}}{a_{3}}$ as well. On the other hand,
the equilibrium compliance is infinite, i.e., $\varepsilon _{cr}^{\left(
e\right) }=\lim_{t\rightarrow \infty }\varepsilon _{cr}\left( t\right)
=\infty $, so that the equilibrium modulus is zero, i.e., $\sigma
_{sr}^{\left( e\right) }=\lim_{t\rightarrow \infty }\sigma _{sr}\left(
t\right) =0$ for both model classes and therefore all fractional Burgers
models describe fluid-like materials. Note, if the equilibrium compliance is
finite, then model would represent the solid-like material.

The implication, proved in the present work, is that fluid-like Burgers
models belonging to the first class have infinite, while the ones belonging
to the second class have finite wave propagation speed%
\begin{equation}
c=\sqrt{\sigma _{sr}^{\left( g\right) }}=\frac{1}{\sqrt{\varepsilon
_{cr}^{\left( g\right) }}}=\sqrt{\frac{b_{2}}{a_{3}}},  \label{ce}
\end{equation}%
as in the case of thermodynamically consistent fractional models arising
from the general fractional linear model 
\begin{equation}
\sum_{i=1}^{n}a_{i}\,{}_{0}\mathrm{D}_{t}^{\alpha _{i}}\sigma
(x,t)=\sum_{j=1}^{m}b_{j}\,{}_{0}\mathrm{D}_{t}^{\beta _{j}}\varepsilon
(x,t),\;\;a_{i},b_{j}>0,\;\alpha _{i},\beta _{j}\in \left( 0,1\right) ,
\label{gen-lin}
\end{equation}%
obtained and analyzed in \cite{AKOZ} for thermodynamical consistency and
used in \cite{KOZ19} as constitutive equations in wave propagation modeling.
Namely, the results of \cite{KOZ10,KOZ11}, where the wave propagation speed
is found via the conic solution support, i.e., $\left\vert x\right\vert <ct,$
in the case of the fractional Zener model and its generalization,
respectively given by%
\begin{gather*}
\left( 1+a\,{}_{0}\mathrm{D}_{t}^{\alpha }\right) \sigma (x,t)=E\left(
1+b\,{}_{0}\mathrm{D}_{t}^{\alpha }\right) \varepsilon (x,t),\;\;0<a\leq
b,\;\alpha \in \left( 0,1\right) , \\
\sum_{i=1}^{n}a_{i}\,{}_{0}\mathrm{D}_{t}^{\alpha _{i}}\sigma
(x,t)=\sum_{i=1}^{n}b_{i}\,{}_{0}\mathrm{D}_{t}^{\alpha _{i}}\varepsilon
(x,t),\;\;0\leq \alpha _{1}\leq \ldots \leq \alpha _{n}<1,\;\frac{a_{1}}{%
b_{1}}\geq \ldots \geq \frac{a_{n}}{b_{n}}\geq 0,
\end{gather*}%
are extended in \cite{KOZ19}, using the same argumentation as in the previous work, to all four classes of thermodynamically consistent linear
fractional models and moreover to the power-type distributed-order model
assuming that the orders of fractional differentiation do not exceed one. In
particular, it is found that both solid-like and fluid-like materials can
have either infinite or finite wave speed. Singularity propagation
properties of the memory and non-local type fractional wave equations are
investigated in \cite{HOZ16,HOZ18} using the tools of microlocal analysis,
supporting the results obtained in \cite{KOZ10}.

Wave propagation phenomena in viscoelastic bodies, modeled by integer and
fractional order models, including the question of wave speed and energy
dissipation properties are analyzed in \cite%
{CaputoMainardi-1971b,CaputoMainardi-1971a}. The wavefront expansion of
solution, due to Buchen and Mainardi, is introduced in \cite{BuchenMainardi}
to be later used in \cite{ColombaroGiustiMainardi2,ColombaroGiustiMainardi1}
when considering the wave equation in viscoelastic materials described by
the Bessel as well as by the integer and fractional order Maxwell and Kelvin-Voigt models.
The Bessel model for viscoelastic body is introduced in \cite{GiustiMainardi} and analyzed in 
\cite{ColombaroGiustiMainardi}. Features of the wave propagation
in viscoelastic media, like the asymptotic behavior of fundamental solution
near the wavefront, dispersion, and attenuation is examined in \cite{Han7,Han8,Han6}. 
Wave propagation speed, reinterpreted as the 
fundamental solution's peak propagation speed is analyzed in \cite%
{LuchkoMainardi-1,LuchkoMainardi-2,LuchkoMainardiPovstenko}.
Modeling viscoelastic materials using the fractional order models, as well as
dispersion and attenuation effects described by the corresponding wave
equations are reviewed in \cite{Mai-10}. 

Fractional wave equations on bounded and semi-bounded domain are considered
in \cite{R-S1,R-S,R-S2} for different fractional models including the Zener,
modified Zener, and modified Maxwell models, as well as in \cite%
{APZ-4,APZ-3,APZ-6} in the case of power-type distributed-order model.
Generalizations of the classical wave equations and corresponding problems
are reviewed in \cite{APSZ-2,R-S-2010}.

\section{Fractional Burgers model in wave propagation}

Fractional Burgers wave equation, as the dimensionless system of equations:%
\begin{equation}
\frac{\partial }{\partial x}\sigma (x,t)=\frac{\partial ^{2}}{\partial t^{2}}%
u(x,t),\;\;\varepsilon (x,t)=\frac{\partial }{\partial x}u(x,t),
\label{em-s}
\end{equation}%
and either%
\begin{equation}
\left( 1+a_{1}\,{}_{0}\mathrm{D}_{t}^{\alpha }+a_{2}\,{}_{0}\mathrm{D}%
_{t}^{\beta }+a_{3}\,{}_{0}\mathrm{D}_{t}^{\gamma }\right) \sigma \left(
x,t\right) =\left( {}_{0}\mathrm{D}_{t}^{\mu }+b\,{}_{0}\mathrm{D}_{t}^{\mu
+\eta }\right) \varepsilon \left( x,t\right)  \label{Burgers1}
\end{equation}%
for the first class of Burgers models, or%
\begin{equation}
\left( 1+a_{1}\,{}_{0}\mathrm{D}_{t}^{\alpha }+a_{2}\,{}_{0}\mathrm{D}%
_{t}^{\beta }+a_{3}\,{}_{0}\mathrm{D}_{t}^{\beta +\eta }\right) \sigma
\left( x,t\right) =\left( {}_{0}\mathrm{D}_{t}^{\beta }+b\,{}_{0}\mathrm{D}%
_{t}^{\beta +\eta }\right) \varepsilon \left( x,t\right) ,  \label{Burgers2}
\end{equation}%
for the second class of Burgers models, subject to initial and boundary
conditions 
\begin{gather}
u(x,0)=u_{0}(x),\;\;\frac{\partial }{\partial t}u(x,0)=v_{0}(x),\;\;\sigma
(x,0)=0,\;\;\varepsilon (x,0)=0,  \label{ic-bd} \\
\lim_{x\rightarrow \pm \infty }u(x,t)=0,\;\;\lim_{x\rightarrow \pm \infty
}\sigma (x,t)=0,  \label{bc-bd}
\end{gather}%
is obtained by introducing dimensionless quantities%
\begin{gather*}
\bar{x}=\frac{x}{\mathcal{U}},\;\;\bar{t}=\frac{t}{T^{\ast }},\;\;\bar{u}=%
\frac{u}{\mathcal{U}},\;\;\bar{u}_{0}=\frac{u_{0}}{\mathcal{U}},\;\;\bar{v}%
_{0}=\frac{T^{\ast }}{\mathcal{U}}v_{0},\;\;\bar{\sigma}=\frac{\sigma }{%
\sigma ^{\ast }},\;\;\bar{a}_{1}=\frac{a_{1}}{\left( T^{\ast }\right)
^{\alpha }},\;\;\bar{a}_{2}=\frac{a_{2}}{\left( T^{\ast }\right) ^{\beta }},
\\
T^{\ast }=\left( \frac{\rho \mathcal{\,U}^{2}}{b_{1}}\right) ^{\frac{1}{2-%
\mathcal{\xi }}},\;\;\sigma ^{\ast }=\left( \frac{b_{1}^{2}}{\left( \rho 
\mathcal{\,U}^{2}\right) ^{\mathcal{\xi }}}\right) ^{\frac{1}{2-\mathcal{\xi 
}}},\;\;\bar{a}_{3}=\frac{a_{3}}{\left( T^{\ast }\right) ^{\zeta }},\;\;\bar{%
b}=\frac{b_{2}}{b_{1}\left( T^{\ast }\right) ^{\eta }},
\end{gather*}%
with $\xi =\mu $ and $\zeta =\gamma $ for the first class of Burgers models, 
$\xi =\beta $ and $\zeta =\beta +\eta $ for the second class, and $\mathcal{U%
}=\sup_{x\in 
\mathbb{R}
}\left\vert u_{0}\left( x\right) \right\vert ,$ into system of governing
equations (\ref{eq-motion}), (\ref{strejn}) and either (\ref{UCE-1-5}) or (%
\ref{UCE-6-8}), subject to (\ref{ic}), (\ref{bc}), and by subsequent
omittance of bars.

Models in dimensionless form, along with the corresponding thermodynamical
restrictions, are listed below.

\noindent \textbf{Model I}: 
\begin{gather}
\left( 1+a_{1}\,{}_{0}\mathrm{D}_{t}^{\alpha }+a_{2}\,{}_{0}\mathrm{D}%
_{t}^{\beta }+a_{3}\,{}_{0}\mathrm{D}_{t}^{\gamma }\right) \sigma \left(
t\right) =\left( {}_{0}\mathrm{D}_{t}^{\mu }+b\,{}_{0}\mathrm{D}_{t}^{\mu
+\eta }\right) \varepsilon \left( t\right) ,  \label{Model 1} \\
0\leq \alpha \leq \beta \leq \gamma \leq \mu \leq 1,\;\;1\leq \mu +\eta \leq
1+\alpha ,\;\;b\leq a_{i}\frac{\cos \frac{\left( \mu -\eta \right) \pi }{2}}{%
\left\vert \cos \frac{\left( \mu +\eta \right) \pi }{2}\right\vert },
\label{Model 1 - tdr}
\end{gather}%
with $\left( \eta ,i\right) \in \left\{ \left( \alpha ,1\right) ,\left(
\beta ,2\right) ,\left( \gamma ,3\right) \right\} ;$

\noindent \textbf{Model II}:%
\begin{gather}
\left( 1+a_{1}\,{}_{0}\mathrm{D}_{t}^{\alpha }+a_{2}\,{}_{0}\mathrm{D}%
_{t}^{\beta }+a_{3}\,{}_{0}\mathrm{D}_{t}^{2\alpha }\right) \sigma \left(
t\right) =\left( {}_{0}\mathrm{D}_{t}^{\mu }+b\,{}_{0}\mathrm{D}_{t}^{\mu
+\alpha }\right) \varepsilon \left( t\right) ,  \label{Model 2} \\
\frac{1}{2}\leq \alpha \leq \beta \leq \mu \leq 1,\;\;\frac{a_{3}}{a_{1}}%
\frac{\left\vert \sin \frac{\left( \mu -2\alpha \right) \pi }{2}\right\vert 
}{\sin \frac{\mu \pi }{2}}\leq b\leq a_{1}\frac{\cos \frac{\left( \mu
-\alpha \right) \pi }{2}}{\left\vert \cos \frac{\left( \mu +\alpha \right)
\pi }{2}\right\vert };  \label{Model 2 - tdr}
\end{gather}

\noindent \textbf{Model III}:%
\begin{gather}
\left( 1+a_{1}\,{}_{0}\mathrm{D}_{t}^{\alpha }+a_{2}\,{}_{0}\mathrm{D}%
_{t}^{\beta }+a_{3}\,{}_{0}\mathrm{D}_{t}^{\alpha +\beta }\right) \sigma
\left( t\right) =\left( {}_{0}\mathrm{D}_{t}^{\mu }+b\,{}_{0}\mathrm{D}%
_{t}^{\mu +\alpha }\right) \varepsilon \left( t\right) ,  \label{Model 3} \\
0\leq \alpha \leq \beta \leq \mu \leq 1,\;\;\alpha +\beta \geq 1,\;\;\frac{%
a_{3}}{a_{2}}\frac{\left\vert \sin \frac{\left( \mu -\beta -\alpha \right)
\pi }{2}\right\vert }{\sin \frac{\left( \mu -\beta +\alpha \right) \pi }{2}}%
\leq b\leq a_{1}\frac{\cos \frac{\left( \mu -\alpha \right) \pi }{2}}{%
\left\vert \cos \frac{\left( \mu +\alpha \right) \pi }{2}\right\vert };
\label{Model 3 - tdr}
\end{gather}

\noindent \textbf{Model IV}:%
\begin{gather}
\left( 1+a_{1}\,{}_{0}\mathrm{D}_{t}^{\alpha }+a_{2}\,{}_{0}\mathrm{D}%
_{t}^{\beta }+a_{3}\,{}_{0}\mathrm{D}_{t}^{\alpha +\beta }\right) \sigma
\left( t\right) =\left( {}_{0}\mathrm{D}_{t}^{\mu }+b\,{}_{0}\mathrm{D}%
_{t}^{\mu +\beta }\right) \varepsilon \left( t\right) ,  \label{Model 4} \\
0\leq \alpha \leq \beta \leq \mu \leq 1,\;\;1-\alpha \leq \beta \leq
1-\left( \mu -\alpha \right) ,\;\;\frac{a_{3}}{a_{1}}\frac{\left\vert \sin 
\frac{\left( \mu -\alpha -\beta \right) \pi }{2}\right\vert }{\sin \frac{%
\left( \mu -\alpha +\beta \right) \pi }{2}}\leq b\leq a_{2}\frac{\cos \frac{%
\left( \mu -\beta \right) \pi }{2}}{\left\vert \cos \frac{\left( \mu +\beta
\right) \pi }{2}\right\vert };  \label{Model 4 - tdr}
\end{gather}

\noindent \textbf{Model V}:%
\begin{gather}
\left( 1+a_{1}\,{}_{0}\mathrm{D}_{t}^{\alpha }+a_{2}\,{}_{0}\mathrm{D}%
_{t}^{\beta }+a_{3}\,{}_{0}\mathrm{D}_{t}^{2\beta }\right) \sigma \left(
t\right) =\left( {}_{0}\mathrm{D}_{t}^{\mu }+b\,{}_{0}\mathrm{D}_{t}^{\mu
+\beta }\right) \varepsilon \left( t\right) ,  \label{Model 5} \\
0\leq \alpha \leq \beta \leq \mu \leq 1,\;\;\frac{1}{2}\leq \beta \leq
1-\left( \mu -\alpha \right) ,\;\;\frac{a_{3}}{a_{2}}\frac{\left\vert \sin 
\frac{\left( \mu -2\beta \right) \pi }{2}\right\vert }{\sin \frac{\mu \pi }{2%
}}\leq b\leq a_{2}\frac{\cos \frac{\left( \mu -\beta \right) \pi }{2}}{%
\left\vert \cos \frac{\left( \mu +\beta \right) \pi }{2}\right\vert }.
\label{Model 5 - tdr}
\end{gather}

\noindent \textbf{Model VI}:%
\begin{gather}
\left( 1+a_{1}\,{}_{0}\mathrm{D}_{t}^{\alpha }+a_{2}\,{}_{0}\mathrm{D}%
_{t}^{\beta }+a_{3}\,{}_{0}\mathrm{D}_{t}^{\alpha +\beta }\right) \sigma
\left( t\right) =\left( {}_{0}\mathrm{D}_{t}^{\beta }+b\,{}_{0}\mathrm{D}%
_{t}^{\alpha +\beta }\right) \varepsilon \left( t\right) ,  \label{Model 6}
\\
0\leq \alpha \leq \beta \leq 1,\;\;\alpha +\beta \geq 1,\;\;\frac{a_{3}}{%
a_{2}}\leq b\leq a_{1}\frac{\cos \frac{\left( \beta -\alpha \right) \pi }{2}%
}{\left\vert \cos \frac{\left( \beta +\alpha \right) \pi }{2}\right\vert };
\label{Model 6 - tdr}
\end{gather}

\noindent \textbf{Model VII}:%
\begin{gather}
\left( 1+a_{1}\,{}_{0}\mathrm{D}_{t}^{\alpha }+a_{2}\,{}_{0}\mathrm{D}%
_{t}^{\beta }+a_{3}\,{}_{0}\mathrm{D}_{t}^{2\beta }\right) \sigma \left(
t\right) =\left( {}_{0}\mathrm{D}_{t}^{\beta }+b\,{}_{0}\mathrm{D}%
_{t}^{2\beta }\right) \varepsilon \left( t\right) ,  \label{Model 7} \\
0\leq \alpha \leq \beta \leq 1,\;\;\frac{1}{2}\leq \beta \leq \frac{1+\alpha 
}{2},\;\;\frac{a_{3}}{a_{2}}\leq b\leq a_{2}\frac{1}{\left\vert \cos \left(
\beta \pi \right) \right\vert };  \label{Model 7 - tdr}
\end{gather}

\noindent \textbf{Model VIII}:%
\begin{gather}
\left( 1+\bar{a}_{1}\,{}_{0}\mathrm{D}_{t}^{\alpha }+\bar{a}_{2}\,{}_{0}%
\mathrm{D}_{t}^{2\alpha }\right) \sigma \left( t\right) =\left( {}_{0}%
\mathrm{D}_{t}^{\alpha }+b\,{}_{0}\mathrm{D}_{t}^{2\alpha }\right)
\varepsilon \left( t\right) ,  \label{Model 8} \\
\frac{1}{2}\leq \alpha \leq 1,\;\;\frac{\bar{a}_{2}}{\bar{a}_{1}}\leq b\leq 
\bar{a}_{1}\frac{1}{\left\vert \cos \left( \alpha \pi \right) \right\vert }.
\label{Model 8 - tdr}
\end{gather}

Application of the Fourier transform with respect to the spatial coordinate%
\begin{equation*}
\hat{f}(\xi )=\mathcal{F}\left[ f\left( x\right) \right] \left( \xi \right)
=\int_{-\infty }^{\infty }f(x)\mathrm{e}^{-\mathrm{i}\xi x}\mathrm{d}%
x,\;\;\xi \in \mathbb{R},
\end{equation*}%
and Laplace transform with respect to the time%
\begin{equation*}
\tilde{f}\left( s\right) =\mathcal{L}\left[ f\left( t\right) \right] \left(
s\right) =\int_{0}^{\infty }f\left( t\right) \mathrm{e}^{-st}\mathrm{d}t,\;\;%
\func{Re}s>0,
\end{equation*}%
with initial (\ref{ic-bd}) and boundary conditions (\ref{bc-bd}) taken into
account, transforms the system of governing equations (\ref{em-s}) and
either (\ref{Burgers1}), or (\ref{Burgers2}) into ($\xi \in \mathbb{R},$ $%
\func{Re}s>0$)%
\begin{gather}
\mathrm{i}\xi \widehat{\tilde{\sigma}}\left( \xi ,s\right) =s^{2}\widehat{%
\tilde{u}}\left( \xi ,s\right) -s\hat{u}_{0}(\xi )+\hat{v}_{0}(\xi ),\;\;%
\widehat{\tilde{\varepsilon}}\left( \xi ,s\right) =\mathrm{i}\xi \widehat{%
\tilde{u}}\left( \xi ,s\right) ,  \label{em-s-ft,lt} \\
\Phi _{\sigma }(s)\widehat{\tilde{\sigma}}\left( \xi ,s\right) =\Phi
_{\varepsilon }(s)\widehat{\tilde{\varepsilon}}\left( \xi ,s\right) ,
\label{fbm-ft,lt}
\end{gather}%
with either%
\begin{equation}
\Phi _{\sigma }(s)=1+a_{1}s^{\alpha }+a_{2}\,s^{\beta }+a_{3}\,s^{\gamma
},\;\;\Phi _{\varepsilon }(s)=s^{\mu }+b\,s^{\mu +\eta },
\label{Burgers1-fiovi}
\end{equation}%
in the case of the first class of Burgers equation (\ref{Burgers1}), or 
\begin{equation}
\Phi _{\sigma }(s)=1+a_{1}s^{\alpha }+a_{2}\,s^{\beta }+a_{3}\,s^{\beta
+\eta },\;\;\Phi _{\varepsilon }(s)=s^{\beta }+b\,s^{\beta +\eta },
\label{Burgers2-fiovi}
\end{equation}%
in the case of the second class of Burgers equation (\ref{Burgers1}).

It is obtained that%
\begin{equation}
\widehat{\tilde{u}}(\xi ,s)=\widehat{\tilde{K}}(\xi ,s)\left( \hat{u}%
_{0}(\xi )+\frac{1}{s}\hat{v}_{0}(\xi )\right) ,\;\;\xi \in \mathbb{R},\;%
\func{Re}s>0,  \label{u-ft,lt}
\end{equation}%
with 
\begin{equation}
\widehat{\tilde{K}}(\xi ,s)=s\frac{\Phi _{\sigma }(s)}{\Phi _{\varepsilon
}(s)}\frac{1}{\xi ^{2}+s^{2}\frac{\Phi _{\sigma }(s)}{\Phi _{\varepsilon }(s)%
}},\;\;\xi \in \mathbb{R},\;\func{Re}s>0,  \label{k-ft,lt}
\end{equation}%
once the system of equations (\ref{em-s-ft,lt}), (\ref{fbm-ft,lt}) is solved
with respect to displacement $\widehat{\tilde{u}}$, implying the solution to
the fractional Burgers equation (\ref{em-s}) and either (\ref{Burgers1}), or
(\ref{Burgers2}), subject to (\ref{ic-bd}) and (\ref{bc-bd}), in the form%
\begin{equation}
u(x,t)=K(x,t)\ast _{x,t}(u_{0}(x)\delta (t)+v_{0}(x)H(t)),
\label{fund sol u}
\end{equation}%
where $\ast _{x}$ denotes the convolution with respect to the spatial
variable: $f\left( x\right) \ast _{x}g\left( x\right) =\int_{-\infty
}^{\infty }f\left( x^{\prime }\right) g\left( x-x^{\prime }\right) \mathrm{d}%
x^{\prime },$ $x\in 
\mathbb{R}
,$ after inverting Fourier and Laplace transforms in (\ref{u-ft,lt}).

In order to calculate the solution kernel $K,$ the inversion of the Fourier
transform is performed in (\ref{k-ft,lt}) using a well-known inversion
formula 
\begin{equation}
\mathcal{F}^{-1}\left[ \frac{1}{\xi ^{2}+\lambda }\right] \left( x\right) =%
\frac{1}{2\sqrt{\lambda }}\mathrm{e}^{-\left\vert x\right\vert \sqrt{\lambda 
}},\;\;x\in 
\mathbb{R}
,\;\lambda \in 
\mathbb{C}
\backslash \left( -\infty ,0\right] ,  \label{FIF}
\end{equation}%
implying 
\begin{equation}
\tilde{K}(x,s)=\frac{1}{2}\sqrt{\frac{\Phi _{\sigma }(s)}{\Phi _{\varepsilon
}(s)}}\mathrm{e}^{-\left\vert x\right\vert s\sqrt{\frac{\Phi _{\sigma }(s)}{%
\Phi _{\varepsilon }(s)}}},\;\;x\in \mathbb{R},\;\func{Re}s>0,  \label{k-lt}
\end{equation}%
provided that 
\begin{equation}
s^{2}\frac{\Phi _{\sigma }(s)}{\Phi _{\varepsilon }(s)}\in \mathbb{C}%
\setminus (-\infty ,0]\;\;\Leftrightarrow \;\;\frac{\Phi _{\sigma }(s)}{\Phi
_{\varepsilon }(s)}\left( s^{2}+\xi ^{2}\frac{\Phi _{\varepsilon }(s)}{\Phi
_{\sigma }(s)}\right) \not=0,\;\;\text{for}\;\;\xi \in 
\mathbb{R}
,\;\func{Re}s>0,  \label{FI-cond}
\end{equation}%
which holds for all Models I - VIII, as proved in Appendix \ref{FIF-ver}.
Further, inverting the Laplace transformation in (\ref{k-lt}) by the
definition 
\begin{equation}
K(x,t)=\mathcal{L}^{-1}\left[ \tilde{K}\left( x,s\right) \right] \left(
t\right) =\frac{1}{2\pi \mathrm{i}}\int_{\Gamma _{0}}\tilde{K}(x,s)\mathrm{e}%
^{st}\mathrm{d}s,\;\;x\in \mathbb{R},\;t>0,  \label{LIF}
\end{equation}%
where $\Gamma _{0}$ is the Bromwich path, the two forms of solution kernel $%
K $ are obtained in Appendix \ref{K-calc} depending on the number and
position of branching points of function $\tilde{K},$ given by (\ref{k-lt}),
originating from the zeros of function $\Phi _{\sigma },$ since $\Phi
_{\varepsilon },$ except for $s=0,$ has no other zeros in the principal
Riemann plane, with $\Phi _{\sigma }$ and $\Phi _{\varepsilon }$ given by
either (\ref{Burgers1-fiovi}) or (\ref{Burgers2-fiovi}). There are three
possible cases, since, as shown in \cite{OZ-2}, function $\Phi _{\sigma }$
can have no zeros, one negative real zero, or a pair of complex conjugated
zeros having negative real part. However, the solution kernel has the same
form in the first two cases, thus merged into Case 1 below, while the form
of the solution kernel differs in the third case, thus being labeled as Case
2.

\textbf{Case 1.} If function $\tilde{K},$ except for $s=0,$ either has no
branching points, or has a negative real branching point, then function $K$
is found as%
\begin{equation}
K\left( x,t\right) =\frac{1}{4\pi \mathrm{i}}\int_{0}^{\infty }\left( \sqrt{%
\frac{\Phi _{\sigma }(\rho \mathrm{e}^{-i\pi })}{\Phi _{\varepsilon }(\rho 
\mathrm{e}^{-i\pi })}}\mathrm{e}^{\left\vert x\right\vert \rho \sqrt{\frac{%
\Phi _{\sigma }(\rho \mathrm{e}^{-i\pi })}{\Phi _{\varepsilon }(\rho \mathrm{%
e}^{-i\pi })}}}-\sqrt{\frac{\Phi _{\sigma }(\rho \mathrm{e}^{i\pi })}{\Phi
_{\varepsilon }(\rho \mathrm{e}^{i\pi })}}\mathrm{e}^{\left\vert
x\right\vert \rho \sqrt{\frac{\Phi _{\sigma }(\rho \mathrm{e}^{i\pi })}{\Phi
_{\varepsilon }(\rho \mathrm{e}^{i\pi })}}}\right) \mathrm{e}^{-\rho t}%
\mathrm{d}\rho ,  \label{fund sol 1}
\end{equation}%
either having support in $%
\mathbb{R}
\times \left[ 0,\infty \right) $ for the first class of fractional Burgers
models, or having support in the conic domain $\left\vert x\right\vert <%
\sqrt{\frac{b}{a_{3}}}t,$ for the second class.

\textbf{Case 2.} If function $\tilde{K},$ except for $s=0,$ has a pair of
complex conjugated branching points with negative real part: $s_{0}=\rho _{0}%
\mathrm{e}^{\mathrm{i}\varphi _{0}}$ and $\bar{s}_{0}=\rho _{0}\mathrm{e}^{-%
\mathrm{i}\varphi _{0}}$, then function $K$ is found as%
\begin{equation}
K\left( x,t\right) \!=\!\frac{1}{4\pi \mathrm{i}}\int_{0}^{\infty
}\!\!\left( \sqrt{\frac{\Phi _{\sigma }(\rho \mathrm{e}^{i\varphi _{0}})}{%
\Phi _{\varepsilon }(\rho \mathrm{e}^{i\varphi _{0}})}}\mathrm{e}^{i\varphi
_{0}-\rho \mathrm{e}^{i\varphi _{0}}\left( \left\vert x\right\vert \sqrt{%
\frac{\Phi _{\sigma }(\rho \mathrm{e}^{i\varphi _{0}})}{\Phi _{\varepsilon
}(\rho \mathrm{e}^{i\varphi _{0}})}}-t\right) }-\sqrt{\frac{\Phi _{\sigma
}(\rho \mathrm{e}^{-i\varphi _{0}})}{\Phi _{\varepsilon }(\rho \mathrm{e}%
^{-i\varphi _{0}})}}\mathrm{e}^{-i\varphi _{0}-\rho \mathrm{e}^{-i\varphi
_{0}}\left( \left\vert x\right\vert \sqrt{\frac{\Phi _{\sigma }(\rho \mathrm{%
e}^{-i\varphi _{0}})}{\Phi _{\varepsilon }(\rho \mathrm{e}^{-i\varphi _{0}})}%
}-t\right) }\right) \!\!\mathrm{d}\rho ,  \label{fund sol 2}
\end{equation}%
either having support in $%
\mathbb{R}
\times \left[ 0,\infty \right) $ for the first class of fractional Burgers
models, or having support in the conic domain $\left\vert x\right\vert <%
\sqrt{\frac{b}{a_{3}}}t,$ for the second class.

The solution support properties, in both cases of solution kernel, define
the wave propagation speed: infinite if the support is $%
\mathbb{R}
\times \left[ 0,\infty \right) ,$ obtained for the first class of Burgers
models, and finite if the support is conic domain $\left\vert x\right\vert <%
\sqrt{\frac{a_{3}}{b}}t,$ obtained as 
\begin{equation}
c=\sqrt{\frac{b}{a_{3}}}  \label{ce1}
\end{equation}%
for the second class of Burgers models. Since $\sigma _{sr}^{\left( g\right)
}=\frac{a_{3}}{b},$ see \cite[Eq. (57)]{OZ-2}, the wave propagation speed (%
\ref{ce1}) is exactly the wave propagation speed (\ref{ce}) that is obtained
in \cite{KOZ19} for the constitutive models having fractional
differentiation orders not exceeding one.

\section{Numerical examples}

\begin{figure}[h]
\begin{center}
\includegraphics[width=0.45\columnwidth]{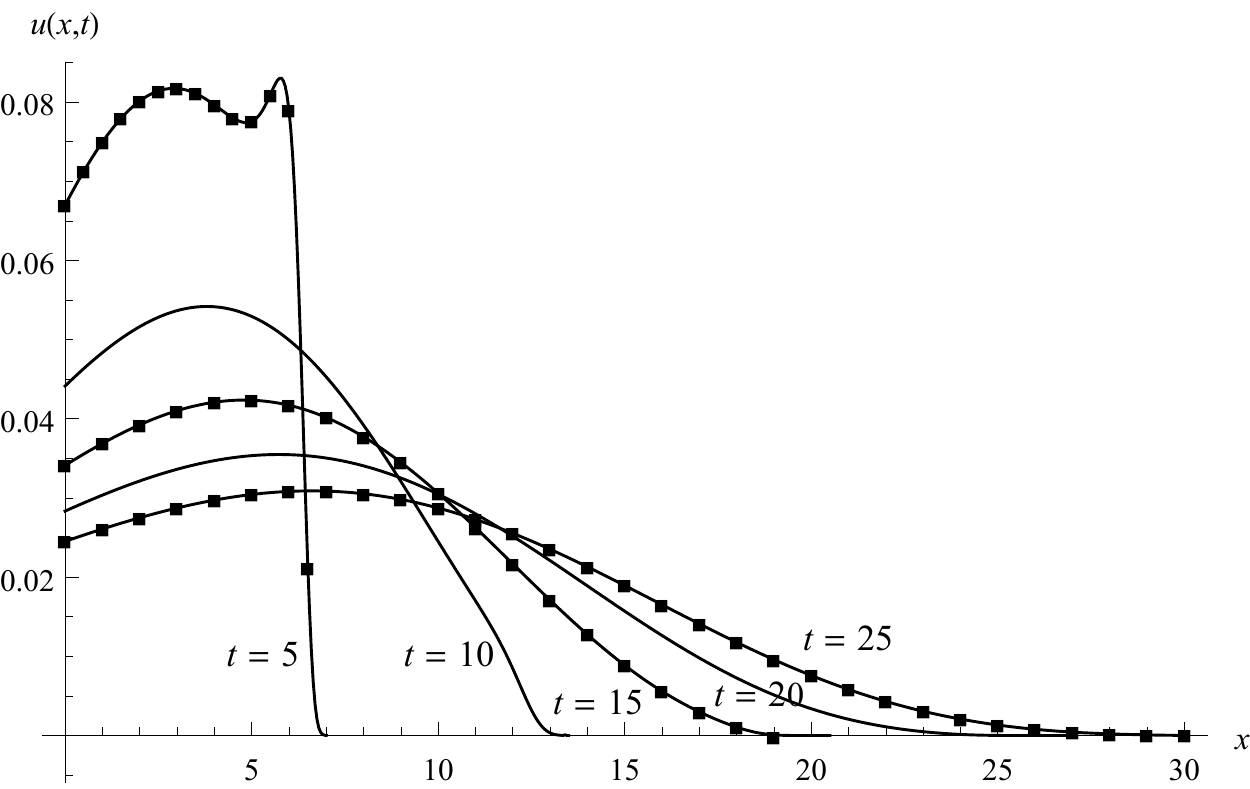}
\end{center}
\caption{Spatial profiles of solution $u$, represented by solid line -
analytical solution, and squares - numerical solution, at different
time-instances for Model V with parameters: $a_{1}=0.075$, $a_{2}=0.8$, $%
a_{3}=1.14$, $b=1.39$, $\protect\alpha =0.4$, $\protect\beta =0.6$, and $%
\protect\mu =0.7$, when, except for $s=0$, there are no other branching
points.}
\label{modV-nema-nula}
\end{figure}

\begin{figure}[p]
\begin{center}
\includegraphics[width=0.45\columnwidth]{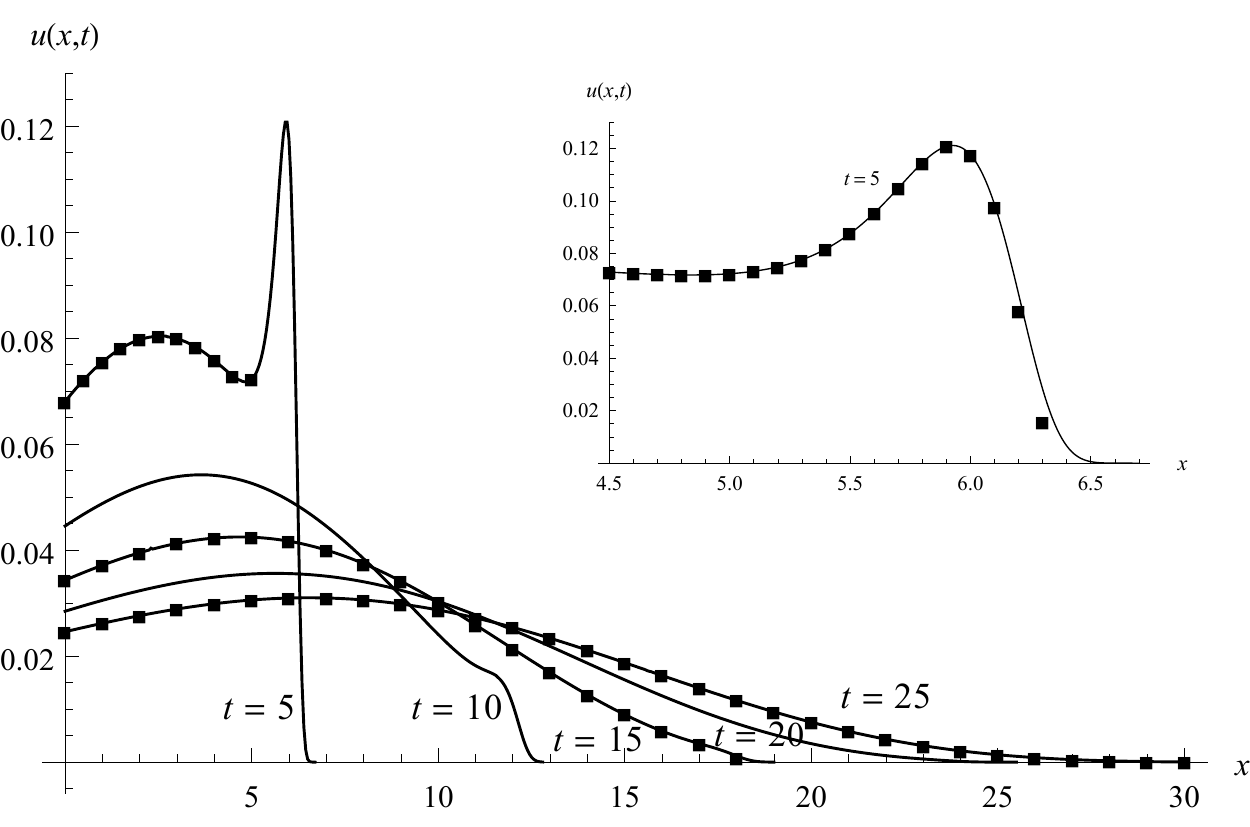}
\end{center}
\caption{Spatial profiles of solution $u$, represented by solid line -
analytical solution, and squares - numerical solution, at different
time-instances for Model V with parameters: $a_{1}=0.075$, $a_{2}=0.8$, $%
a_{3}=1.14$, $b=1.39$, $\protect\alpha =0.4$, $\protect\beta =0.63138$, and $%
\protect\mu =0.7$, when, except for $s=0$, there is one real branching
point. }
\label{modV-real-nule}
\end{figure}

\begin{figure}[p]
\begin{center}
\begin{minipage}{0.45\columnwidth}
   \subfloat[$\beta=0.7$ - no branching points]{
   \includegraphics[width=0.7\columnwidth]{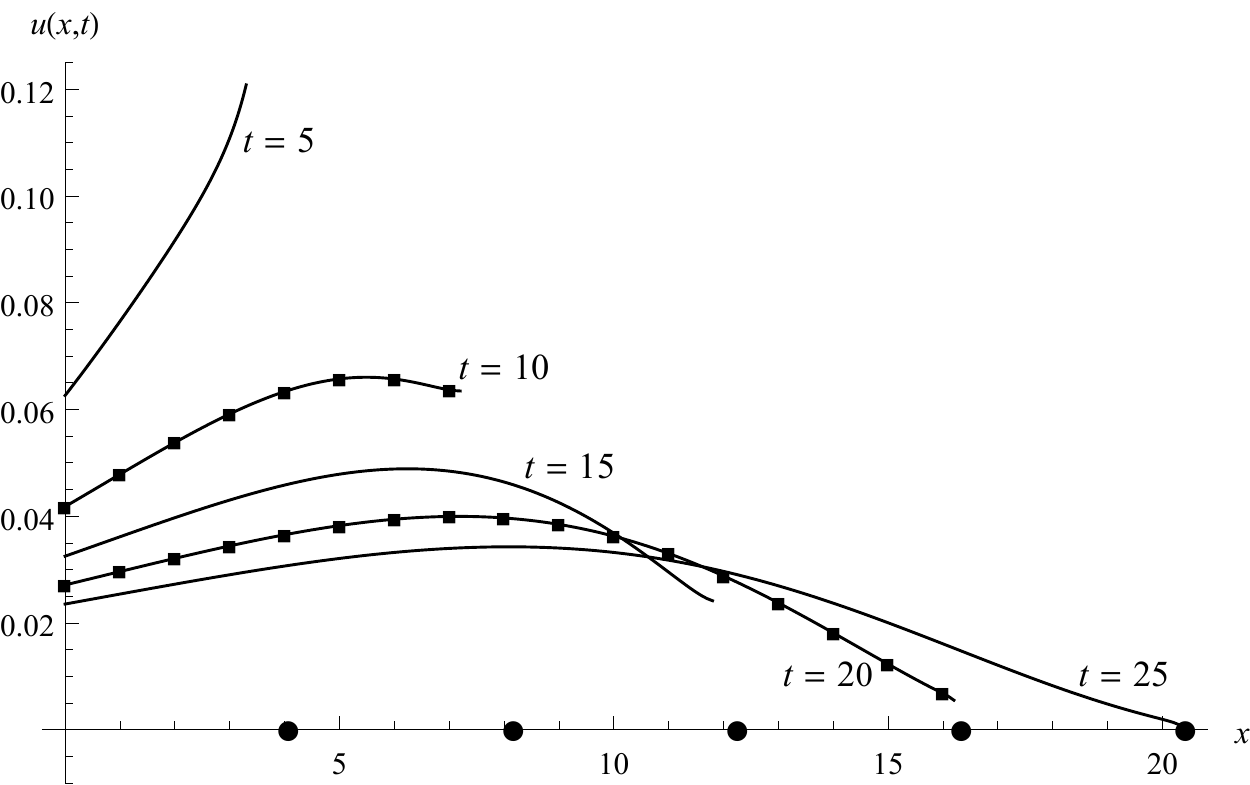}
   \label{m7-n-n}}
  \end{minipage}
\smallskip \vfil
\begin{minipage}{0.45\columnwidth}
   \subfloat[$\beta=0.76976$ - one real branching point]{
   \includegraphics[width=0.7\columnwidth]{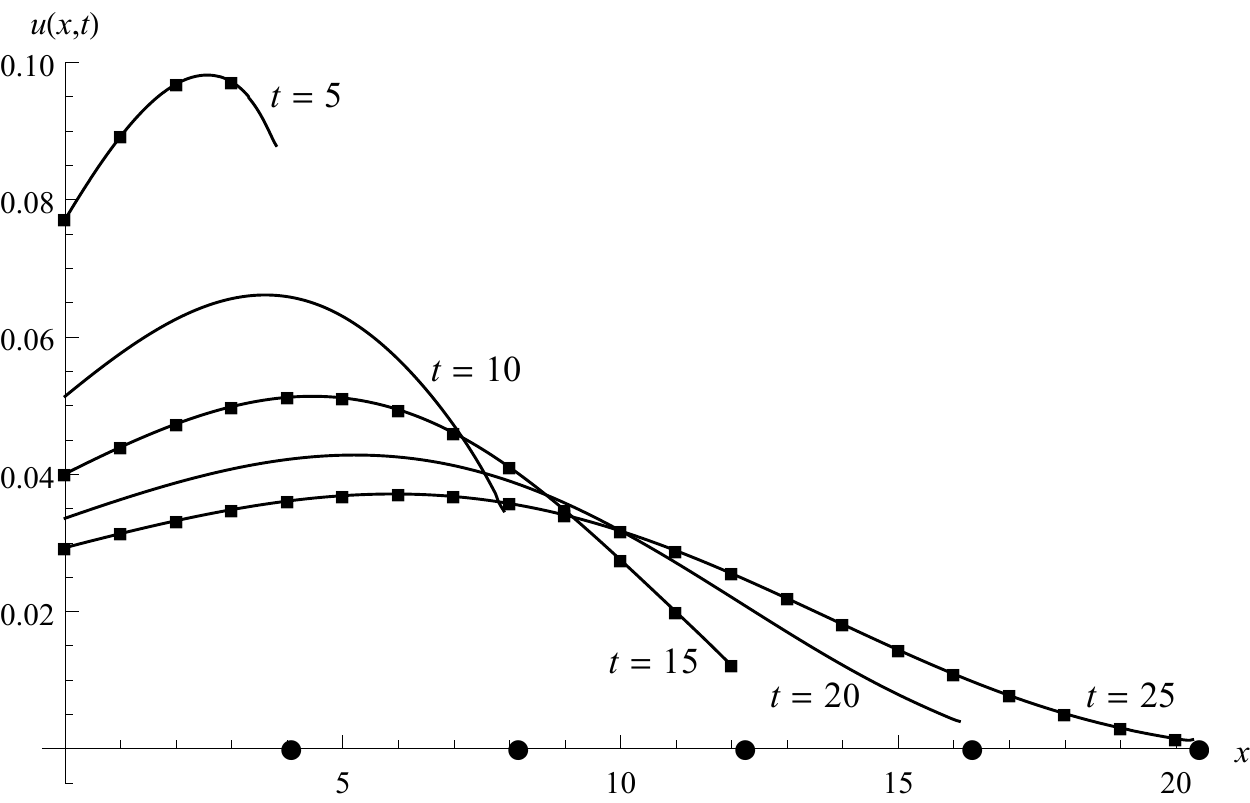}
   \label{m7-r-n}}
  \end{minipage}
\hfil
\begin{minipage}{0.45\columnwidth}
   \subfloat[$\beta=0.79$ - pair of complex conjugated branching points]{
   \includegraphics[width=0.7\columnwidth]{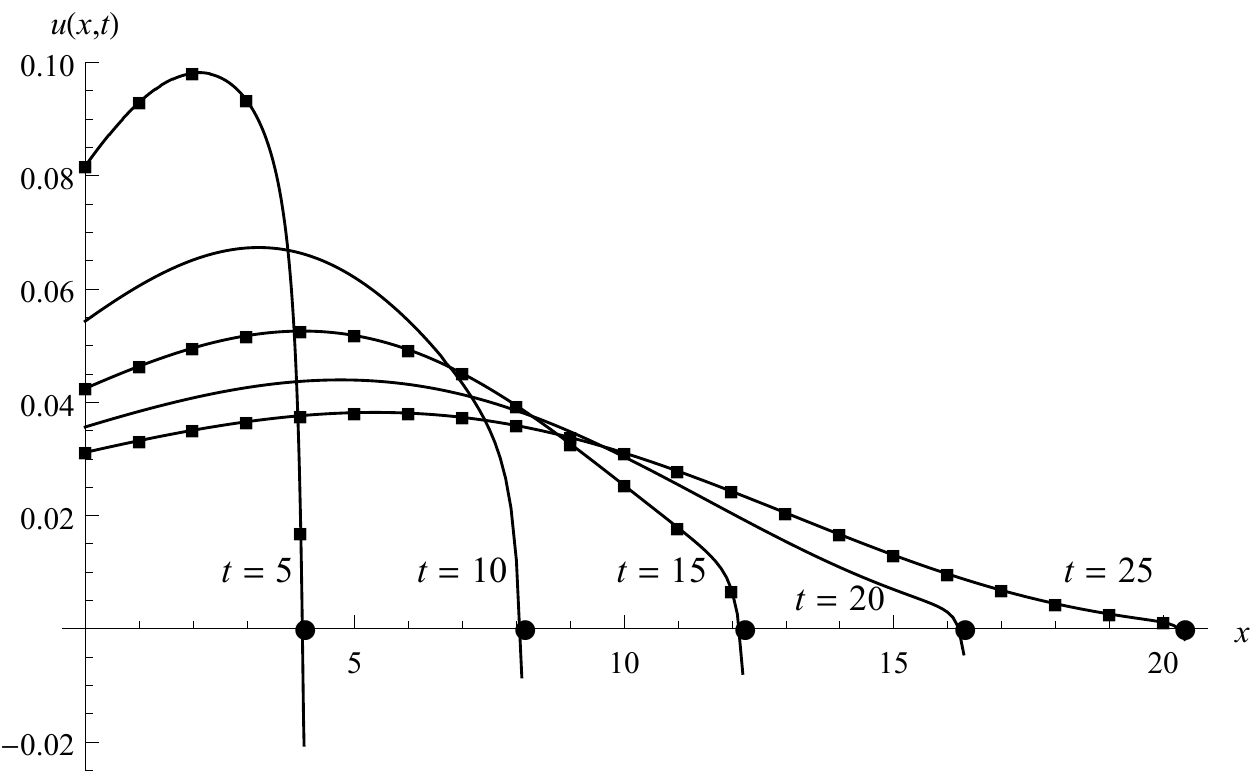}
   \label{m7-k-n}}
  \end{minipage}
\end{center}
\caption{Spatial profiles of solution $u$, represented by solid line -
analytical solution, and squares - numerical solution, while circles
represent ending points of solution support, at different time-instances for
Model VII with parameters: $a_{1}=1.25$, $a_{2}=1.5$, $a_{3}=2.825$, $%
b=1.885 $, and $\protect\alpha =0.6$. There are three cases corresponding to
different number of branching points, except $s=0$, depending on $\protect%
\beta $.}
\label{modVII-isti}
\end{figure}

\begin{figure}[h]
\begin{center}
\includegraphics[width=0.45\columnwidth]{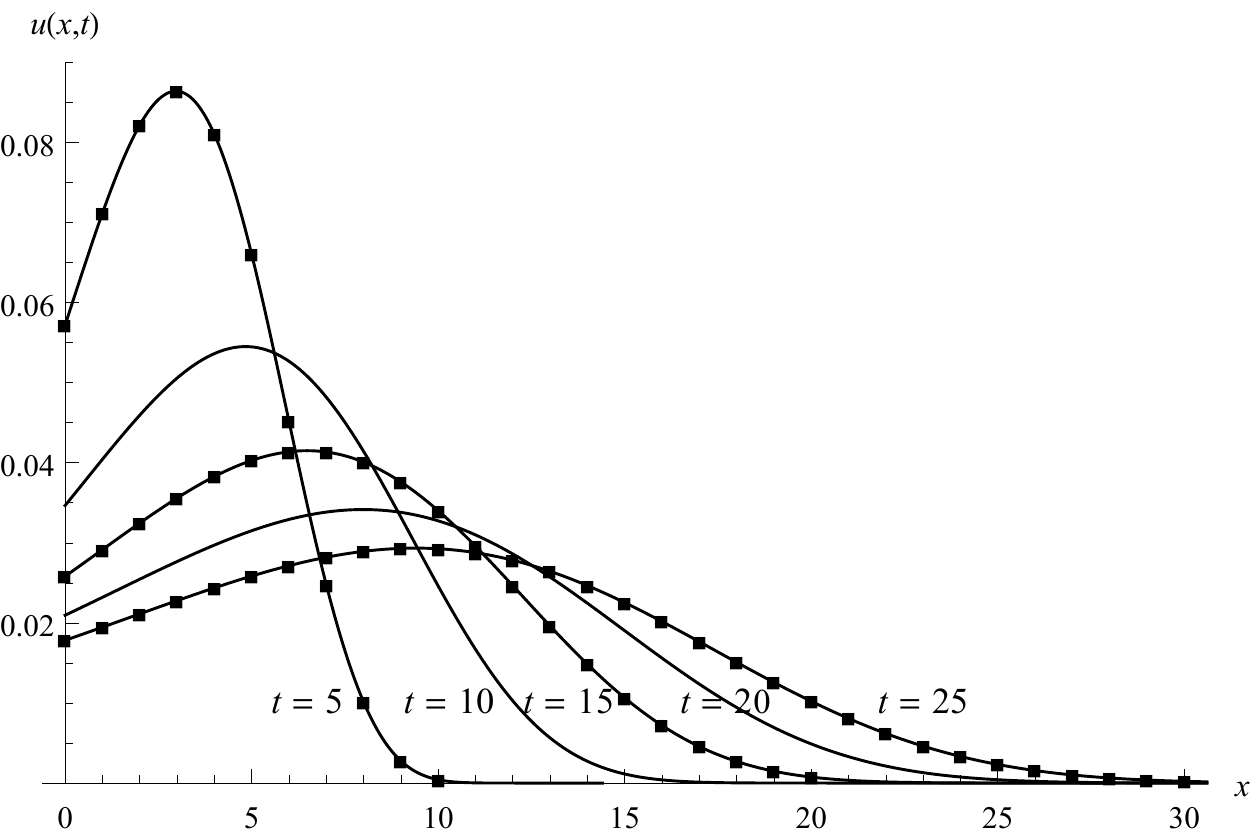}
\end{center}
\caption{Spatial profiles of solution $u$, represented by solid line -
analytical solution, and squares - numerical solution, at different
time-instances for Model VII with parameters: $a_{1}=0.25$, $a_{2}=0.75$, $%
a_{3}=0.15$, $b=1.25$, $\protect\alpha =0.2$, and $\protect\beta =0.59$,
when, except for $s=0$, there are no other branching points.}
\label{modVII-nema-nule}
\end{figure}

\begin{figure}[h]
\begin{center}
\includegraphics[width=0.45\columnwidth]{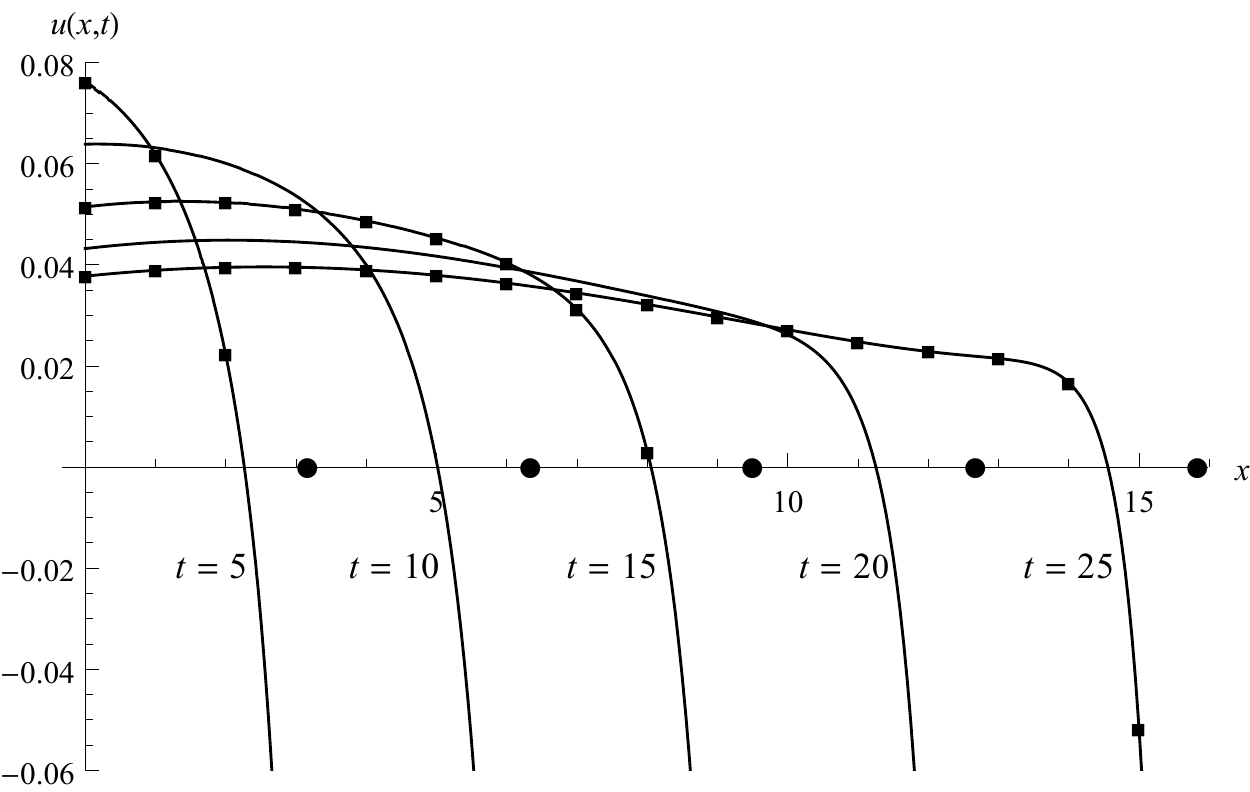}
\end{center}
\caption{Spatial profiles of solution $u$, represented by solid line -
analytical solution, and squares - numerical solution, while circles
represent ending points of solution support, at different time-instances for
Model VII with parameters: $a_{1}=0.01$, $a_{2}=2.5$, $a_{3}=7$, $b=2.81$, $%
\protect\alpha =0.7$, and $\protect\beta =0.845$, when, except for $s=0$,
there is a pair of complex conjugated branching points.}
\label{modVII-kompl-nule}
\end{figure}

Spatial profiles of the solution to the fractional Burgers wave equations,
written as the system of equations (\ref{em-s}) and either (\ref{Burgers1}),
or (\ref{Burgers2}), subject to initial and boundary conditions (\ref{ic-bd}%
) and (\ref{bc-bd}), with the initial displacement being the Dirac delta
distribution and initial velocity being zero, i.e., $u_{0}=\delta ,$ and $%
v_{0}=0,$ implying that the solution is equal to the solution kernel $K,$
are depicted in Figures \ref{modV-nema-nula}, \ref{modV-real-nule}, and \ref%
{modV-kompl-nule} for Model V, representing the first class of fractional
Burgers models and in Figures \ref{modVII-isti}, \ref{modVII-nema-nule}, and %
\ref{modVII-kompl-nule} for Model VII, representing the second class.
Recall, in the case of constitutive models belonging to the first class the
wave propagation speed is infinite, while in the case of the second class
the speed is finite and given by (\ref{ce}). Spatial profiles produced by
using the analytical formula for solution kernel $K,$ given by either (\ref%
{fund sol 1}), or (\ref{fund sol 2}), are compared with the solution kernel
numerically calculated by the fixed Talbot numerical Laplace inversion 
\textit{Mathematica} function, developed by J. Abate and P. P. Valk\'{o}
according to \cite{AbateValko} and available at:
http://library.wolfram.com/infocenter/MathSource/4738/. In each of the
numerical examples good agreement between profiles obtained by these two
methods is found.

\begin{figure}[p]
\begin{center}
\includegraphics[width=0.45\columnwidth]{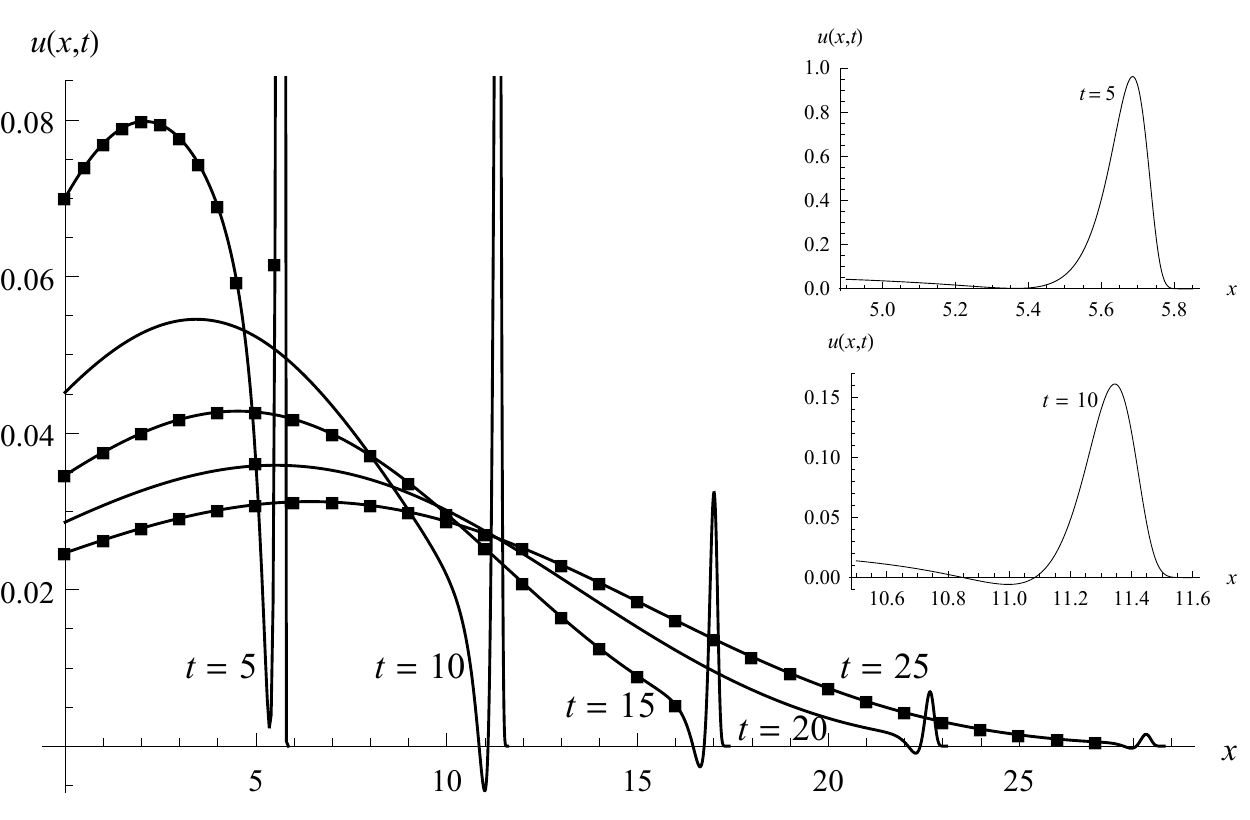}
\end{center}
\caption{Spatial profiles of solution $u$, represented by solid line -
analytical solution, and squares - numerical solution, at different
time-instances for Model V with parameters: $a_{1}=0.075$, $a_{2}=0.8$, $%
a_{3}=1.14$, $b=1.39$, $\protect\alpha =0.4$, $\protect\beta =0.685$, and $%
\protect\mu =0.7$, when, except for $s=0$, there is a pair of complex
conjugated branching points.}
\label{modV-kompl-nule}
\end{figure}

Figures \ref{modV-nema-nula}, \ref{modV-real-nule}, and \ref{modV-kompl-nule}
present spatial profiles for Model V in cases when function $\tilde{K},$
given by (\ref{k-lt}), except for $s=0$ does not have other branching
points, has one negative real, and has a pair of complex conjugated
branching points, respectively. Different number and position of the
branching points is a consequence of the change of a single parameter $\beta
.$ Apart from the main peak originating from the propagation of the initial
Dirac delta displacement, there is a noticeable additional peak that is more
prominent for small times and ceasing as time passes. As the parameter $%
\beta $ increases, the change of the nature (number and position) of the
branching points from no branching points to a pair of complex conjugated
ones, implies the growth of prominence of the additional peak. During the
propagation, due to the energy dissipation, height of the main peak
decreases, while the width of profile is increasing, while propagation
itself is rather slow. 

Wave propagation speed is finite for the second class of fractional Burgers
models, and in Figures \ref{modVII-isti}, \ref{modVII-nema-nule}, and \ref%
{modVII-kompl-nule}, presenting spatial profiles for Model VII, it is
underlined by denoting the ending points of solution support by circles. It
is also noticeable that during the propagation, due to the energy
dissipation, height of the peak decreases, while its width increases.

Figure \ref{modVII-isti} presents spatial profiles depending on the nature
of the branching points, different than $s=0,$ of function $\tilde{K}$ given
by (\ref{k-lt}) in three cases obtained as a consequence of changing
parameter $\beta $: Figure \ref{m7-n-n} represents case when there are no
other branching points, Figure \ref{m7-r-n} when there is one negative real
branching point, and Figure \ref{m7-k-n} when there is a pair of complex
conjugated branching points. For small times, the profile shapes are
considerably different, while as time passes the profile shapes become
alike. In all cases there are jumps at the ending points of solution
support: in Figures \ref{m7-n-n} and \ref{m7-r-n} displacement jumps from a
positive value to zero, while in Figure \ref{m7-k-n} displacement jumps from
a negative value to zero.

When compared to the profiles from Figure \ref{m7-n-n}, where the
displacement jumps to zero at the ending point of solution support, the
displacements plotted in Figure \ref{modVII-nema-nule}, representing also
the case when there are no other branching points than $s=0,$ tend smoothly
to zero at the ending points of solution support. Profiles from Figure \ref%
{modVII-nema-nule} are similar to the profiles obtained in \cite%
{KOZ10,KOZ11,KOZ19} for fractional constitutive models used wave propagation
modeling in viscoelastic dissipative media. 

Figure \ref{modVII-kompl-nule} presents spatial profiles in another case of
model parameters yielding existence of a pair of complex conjugated
branching points (apart of $s=0$) which differ from the ones presented in
Figure \ref{m7-k-n}, since it seems that peaks are situated at zero, while
displacement seems to converge to infinity at the ending point of solution
support. 

\section{Conclusion}

Fractional Burgers wave equations, considered as a dimensionless system of:
equation of motion and strain (\ref{em-s}), coupled with the constitutive
Burgers models either of the first class (\ref{Burgers1}), or of the second
class (\ref{Burgers2}), are solved for the Cauchy initial value problem and
their solutions as a response to the initial Dirac delta displacement with
zero initial velocity are qualitatively analyzed through numerical examples.
The method of Fourier, with respect to space, and Laplace transform with
respect to time are used in order to obtain analytical solution as a
convolution of the solution kernels and initial data. The form of the
solution kernel proved to be dependant on model parameters, so that if
parameters yield, except for $s=0,$ either no branching points, or one
negative real branching point of the Laplace transform of solution kernel,
then solution kernel takes the form (\ref{fund sol 1}), while if, except for 
$s=0,$ the Laplace transform of solution kernel has a pair of complex
conjugated branching points, then solution kernel takes the form (\ref{fund
sol 2}).

Arising from the solution support properties, in both cases of solution
kernel, the infinite wave propagation speed is obtained for the first class
of Burgers models and finite for the second class. Moreover, the obtained
wave propagation speed is consistent with the one obtained for the wave
equations involving fractional linear models with differentiation orders
below one.

Qualitative analysis has shown the dissipative behavior for both classes of
Burgers wave equations, as expected from thermodynamically consistent
constitutive laws for viscoelastic body. However, spatial profile shapes
differs for the different nature of the branching points. The features of
spatial profiles include the jumps from finite value of displacement to zero
at the ending points of solution support, as well as profiles that are not
expected in wave propagation behavior, like occurrence of the additional
peaks and peaks situated at zero.

\appendix

\section{Justification for using the Fourier inversion formula \label%
{FIF-ver}}

The solution kernel is obtained by the Fourier and Laplace transforms as (%
\ref{k-ft,lt}), and in order to apply the Fourier transform inversion
formula (\ref{FIF}), the condition (\ref{FI-cond}), i.e., 
\begin{equation*}
\frac{\Phi _{\sigma }(s)}{\Phi _{\varepsilon }(s)}\left( s^{2}+\xi ^{2}\frac{%
\Phi _{\varepsilon }(s)}{\Phi _{\sigma }(s)}\right) \not=0,\;\;\text{for}%
\;\;\xi \in 
\mathbb{R}
,\;\func{Re}s>0,
\end{equation*}%
must be fulfilled.

Functions $\Phi _{\sigma }$ and $\Phi _{\varepsilon },$ given by (\ref%
{Burgers1-fiovi}) in the case of the first, or by (\ref{Burgers2-fiovi}) in
the case of the second model class, are never zero for $\func{Re}s>0.$
Namely, it is well-known that function $\Phi _{\varepsilon },$ except for $%
s=0,$ does not have other zeros in the principal Riemann branch $\arg s\in
\left( -\pi ,\pi \right) ,$ while for function $\Phi _{\sigma }$ it is
proved in \cite{OZ-2} that if it has zeros, then they lie in the left
complex half-plane.

Therefore, it is left to prove that%
\begin{equation}
\psi \left( s\right) =s^{2}+\xi ^{2}\frac{\Phi _{\varepsilon }(s)}{\Phi
_{\sigma }(s)}\not=0,\;\;\text{for}\;\;\xi \in 
\mathbb{R}
,\;\func{Re}s>0.  \label{malo-psi}
\end{equation}%
It is clear that if $s=\rho >0,$ then%
\begin{equation*}
\psi \left( \rho \right) =\rho ^{2}+\xi ^{2}\rho ^{\mu }\frac{1+b\,\rho
^{\eta }}{1+a_{1}\rho ^{\alpha }+a_{2}\,\rho ^{\beta }+a_{3}\,\rho ^{\gamma }%
}>0.
\end{equation*}%
Further, by substituting $s=\rho \mathrm{e}^{\mathrm{i}\varphi },$ $\varphi
\in \left( -\frac{\pi }{2},\frac{\pi }{2}\right) ,$ into (\ref{malo-psi})
one obtains%
\begin{equation*}
\func{Im}\psi \left( \rho ,\varphi \right) =\rho ^{2}\sin \left( 2\varphi
\right) +\frac{\xi ^{2}\rho ^{\mu }}{\left\vert \Phi _{\sigma }\left( \rho
,\varphi \right) \right\vert ^{2}}f_{\rho }\left( \varphi \right) ,
\end{equation*}%
with%
\begin{align}
f_{\rho }\left( \varphi \right) & =\sin \left( \mu \varphi \right) +b\rho
^{\eta }\sin \left( \left( \mu +\eta \right) \varphi \right) +a_{1}\rho
^{\alpha }\sin \left( \left( \mu -\alpha \right) \varphi \right) +a_{1}b\rho
^{\alpha +\eta }\sin \left( \left( \mu +\eta -\alpha \right) \varphi \right)
\notag \\
& +a_{2}\rho ^{\beta }\sin \left( \left( \mu -\beta \right) \varphi \right)
+a_{2}b\rho ^{\beta +\eta }\sin \left( \left( \mu +\eta -\beta \right)
\varphi \right) +a_{3}\rho ^{\gamma }\sin \left( \left( \mu -\gamma \right)
\varphi \right) +a_{3}b\rho ^{\gamma +\eta }\sin \left( \left( \mu +\eta
-\gamma \right) \varphi \right) ,  \label{f-od-fi}
\end{align}%
that will for each fractional Burgers model prove to be strictly positive if 
$\varphi \in \left( 0,\frac{\pi }{2}\right) $ implying that $\psi ,$ given
by (\ref{malo-psi}) cannot be zero for $\func{Re}s>0.$ Since $\func{Im}\psi
\left( \rho ,-\varphi \right) =-\func{Im}\psi \left( \rho ,\varphi \right) ,$
note that $\func{Im}\psi \left( \rho ,\varphi \right) <0$ if $\varphi \in
\left( -\frac{\pi }{2},0\right) .$

\paragraph{Model I}

is obtained for $\eta \in \left\{ \alpha ,\beta ,\gamma \right\} ,$ so that
function $f_{\rho },$ given by (\ref{f-od-fi}), reads%
\begin{align}
f_{\rho }\left( \varphi \right) & =\sin \left( \mu \varphi \right)
+a_{1}\rho ^{\alpha }\sin \left( \left( \mu -\alpha \right) \varphi \right)
+a_{2}\rho ^{\beta }\sin \left( \left( \mu -\beta \right) \varphi \right)
+a_{3}\rho ^{\gamma }\sin \left( \left( \mu -\gamma \right) \varphi \right) 
\notag \\
& +\left\{ 
\begin{tabular}{l}
$b\rho ^{\alpha }\sin \left( \left( \mu +\alpha \right) \varphi \right)
+a_{1}b\rho ^{2\alpha }\sin \left( \mu \varphi \right) +a_{2}b\rho ^{\alpha
+\beta }\sin \left( \left( \mu -\beta +\alpha \right) \varphi \right)
+a_{3}b\rho ^{\alpha +\gamma }\sin \left( \left( \mu -\gamma +\alpha \right)
\varphi \right) ,$\medskip \\ 
$b\rho ^{\beta }\sin \left( \left( \mu +\beta \right) \varphi \right)
+a_{1}b\rho ^{\alpha +\beta }\sin \left( \left( \mu -\alpha +\beta \right)
\varphi \right) +a_{2}b\rho ^{2\beta }\sin \left( \mu \varphi \right)
+a_{3}b\rho ^{\gamma +\beta }\sin \left( \left( \mu -\gamma +\beta \right)
\varphi \right) ,$\medskip \\ 
$b\rho ^{\gamma }\sin \left( \left( \mu +\gamma \right) \varphi \right)
+a_{1}b\rho ^{\alpha +\gamma }\sin \left( \left( \mu -\alpha +\gamma \right)
\varphi \right) +a_{2}b\rho ^{\beta +\gamma }\sin \left( \left( \mu -\beta
+\gamma \right) \varphi \right) +a_{3}b\rho ^{2\gamma }\sin \left( \mu
\varphi \right) .$%
\end{tabular}%
\right.  \label{f-model-1}
\end{align}%
The thermodynamical restrictions (\ref{Model 1 - tdr}) imply the positivity
of all terms in (\ref{f-model-1}), yielding $f_{\rho }\left( \varphi \right)
>0$ if $\varphi \in \left( 0,\frac{\pi }{2}\right) .$

\paragraph{Model II}

is obtained for $\gamma =2\alpha $ and $\eta =\alpha ,$ so that function $%
f_{\rho },$ given by (\ref{f-od-fi}), reads%
\begin{align}
f_{\rho }\left( \varphi \right) & =\sin \left( \mu \varphi \right) +b\rho
^{\alpha }\sin \left( \left( \mu +\alpha \right) \varphi \right) +a_{1}\rho
^{\alpha }\sin \left( \left( \mu -\alpha \right) \varphi \right) +a_{2}\rho
^{\beta }\sin \left( \left( \mu -\beta \right) \varphi \right)  \notag \\
& +a_{2}b\rho ^{\alpha +\beta }\sin \left( \left( \mu -\beta +\alpha \right)
\varphi \right) +a_{3}b\rho ^{3\alpha }\sin \left( \left( \mu -\alpha
\right) \varphi \right) +a_{1}\rho ^{2\alpha }\sin \left( \mu \varphi
\right) \left( b-\frac{a_{3}}{a_{1}}\frac{\left\vert \sin \left( \left( \mu
-2\alpha \right) \varphi \right) \right\vert }{\sin \left( \mu \varphi
\right) }\right) .  \label{f-model-2}
\end{align}

Consider function $g$ and its first derivative $g^{\prime }$:%
\begin{equation}
g\left( \varphi \right) =\frac{\sin \left( \zeta \varphi \right) }{\sin
\left( \xi \varphi \right) }\;\;\text{and}\;\;g^{\prime }\left( \varphi
\right) =\frac{\xi \varphi \,\zeta \varphi \,\cos \left( \xi \varphi \right)
\cos \left( \zeta \varphi \right) }{\varphi \sin ^{2}\left( \xi \varphi
\right) }\left( \frac{\tan \left( \xi \varphi \right) }{\xi \varphi }-\frac{%
\tan \left( \zeta \varphi \right) }{\zeta \varphi }\right) ,  \label{g}
\end{equation}%
on the interval $\varphi \in \left( 0,\frac{\pi }{2}\right) .$ Let $0<\zeta
<\xi <1.$ Since function $\frac{\tan x}{x}$ is monotonically increasing for $%
x\in \left( 0,\frac{\pi }{2}\right) ,$ one has $g^{\prime }\left( \varphi
\right) >0,$ $\varphi \in \left( 0,\frac{\pi }{2}\right) ,$ implying that
function $g$ is an increasing function on the same interval and therefore%
\begin{equation}
g\left( \varphi \right) <g\left( \frac{\pi }{2}\right) ,\;\;\text{for}%
\;\;\varphi \in \left( 0,\frac{\pi }{2}\right) .  \label{g-manje-g}
\end{equation}

The thermodynamical restriction (\ref{Model 2 - tdr}) yields $0<2\alpha -\mu
<\mu <1,$ so that by setting $\zeta =2\alpha -\mu $ and $\xi =\mu $ in
function $g$ given by (\ref{g}), using (\ref{g-manje-g}) one has 
\begin{equation*}
\frac{\sin \left( \left( 2\alpha -\mu \right) \varphi \right) }{\sin \left(
\mu \varphi \right) }<\frac{\left\vert \sin \frac{\left( \mu -2\alpha
\right) \pi }{2}\right\vert }{\sin \frac{\mu \pi }{2}}.
\end{equation*}%
Therefore, again by (\ref{Model 2 - tdr}), one has that $b-\frac{a_{3}}{a_{1}%
}\frac{\left\vert \sin \left( \left( \mu -2\alpha \right) \varphi \right)
\right\vert }{\sin \left( \mu \varphi \right) }>0,$ which, along with the
positivity of all other terms in (\ref{f-model-2}), implies that $f_{\rho
}\left( \varphi \right) >0$ if $\varphi \in \left( 0,\frac{\pi }{2}\right) .$

\paragraph{Model III}

is obtained for $\gamma =\alpha +\beta $ and $\eta =\alpha ,$ so that
function $f_{\rho },$ given by (\ref{f-od-fi}), reads%
\begin{align}
f_{\rho }\left( \varphi \right) & =\sin \left( \mu \varphi \right) +b\rho
^{\alpha }\sin \left( \left( \mu +\alpha \right) \varphi \right) +a_{1}\rho
^{\alpha }\sin \left( \left( \mu -\alpha \right) \varphi \right) +a_{1}b\rho
^{2\alpha }\sin \left( \mu \varphi \right) +a_{2}\rho ^{\beta }\sin \left(
\left( \mu -\beta \right) \varphi \right)  \notag \\
& +a_{3}b\rho ^{2\alpha +\beta }\sin \left( \left( \mu -\beta \right)
\varphi \right) +a_{2}\rho ^{\alpha +\beta }\sin \left( \left( \mu -\beta
+\alpha \right) \varphi \right) \left( b-\frac{a_{3}}{a_{2}}\frac{\left\vert
\sin \left( \left( \mu -\beta -\alpha \right) \varphi \right) \right\vert }{%
\sin \left( \left( \mu -\beta +\alpha \right) \varphi \right) }\right) .
\label{f-model-3}
\end{align}%
The thermodynamical restriction (\ref{Model 3 - tdr}) yields $0<\alpha
-\left( \mu -\beta \right) <\alpha +\left( \mu -\beta \right) <1,$ so that
by setting $\zeta =\alpha -\left( \mu -\beta \right) $ and $\xi =\alpha
+\left( \mu -\beta \right) $ in function $g$ given by (\ref{g}), using (\ref%
{g-manje-g}) one has 
\begin{equation*}
\frac{\sin \left( \left( \alpha +\beta -\mu \right) \varphi \right) }{\sin
\left( \left( \mu -\beta +\alpha \right) \varphi \right) }<\frac{\left\vert
\sin \frac{\left( \mu -\beta -\alpha \right) \pi }{2}\right\vert }{\sin 
\frac{\left( \mu -\beta +\alpha \right) \pi }{2}}.
\end{equation*}%
Therefore, again by (\ref{Model 3 - tdr}), one has that $b-\frac{a_{3}}{a_{1}%
}\frac{\left\vert \sin \left( \left( \mu -\beta -\alpha \right) \varphi
\right) \right\vert }{\sin \left( \left( \mu -\beta +\alpha \right) \varphi
\right) }>0,$ which, along with the positivity of all other terms in (\ref%
{f-model-3}), implies that $f_{\rho }\left( \varphi \right) >0$ if $\varphi
\in \left( 0,\frac{\pi }{2}\right) .$

\paragraph{Model IV}

is obtained for $\gamma =\alpha +\beta $ and $\eta =\beta ,$ so that
function $f_{\rho },$ given by (\ref{f-od-fi}), reads%
\begin{align}
f_{\rho }\left( \varphi \right) & =\sin \left( \mu \varphi \right) +b\rho
^{\beta }\sin \left( \left( \mu +\beta \right) \varphi \right) +a_{1}\rho
^{\alpha }\sin \left( \left( \mu -\alpha \right) \varphi \right) +a_{2}\rho
^{\beta }\sin \left( \left( \mu -\beta \right) \varphi \right) +a_{2}b\rho
^{2\beta }\sin \left( \mu \varphi \right)  \notag \\
& +a_{3}b\rho ^{\alpha +2\beta }\sin \left( \left( \mu -\alpha \right)
\varphi \right) +a_{1}\rho ^{\alpha +\beta }\sin \left( \left( \mu -\alpha
+\beta \right) \varphi \right) \left( b-\frac{a_{3}}{a_{1}}\frac{\left\vert
\sin \left( \left( \mu -\alpha -\beta \right) \varphi \right) \right\vert }{%
\sin \left( \left( \mu -\alpha +\beta \right) \varphi \right) }\right) .
\label{f-model-4}
\end{align}%
The thermodynamical restriction (\ref{Model 4 - tdr}) yields $0<\beta
-\left( \mu -\alpha \right) <\beta +\left( \mu -\alpha \right) <1,$ so that
by setting $\zeta =\beta -\left( \mu -\alpha \right) $ and $\xi =\beta
+\left( \mu -\alpha \right) $ in function $g$ given by (\ref{g}), using (\ref%
{g-manje-g}) one has 
\begin{equation*}
\frac{\sin \left( \left( \alpha +\beta -\mu \right) \varphi \right) }{\sin
\left( \left( \mu -\alpha +\beta \right) \varphi \right) }<\frac{\left\vert
\sin \frac{\left( \mu -\alpha -\beta \right) \pi }{2}\right\vert }{\sin 
\frac{\left( \mu -\alpha +\beta \right) \pi }{2}}.
\end{equation*}%
Therefore, again by (\ref{Model 4 - tdr}), one has that $b-\frac{a_{3}}{a_{1}%
}\frac{\left\vert \sin \left( \left( \mu -\alpha -\beta \right) \varphi
\right) \right\vert }{\sin \left( \left( \mu -\alpha +\beta \right) \varphi
\right) }>0,$ which, along with the positivity of all other terms in (\ref%
{f-model-4}), implies that $f_{\rho }\left( \varphi \right) >0$ if $\varphi
\in \left( 0,\frac{\pi }{2}\right) .$

\paragraph{Model V}

is obtained for $\gamma =2\beta $ and $\eta =\beta ,$ so that function $%
f_{\rho },$ given by (\ref{f-od-fi}), reads%
\begin{align}
f_{\rho }\left( \varphi \right) & =\sin \left( \mu \varphi \right) +b\rho
^{\beta }\sin \left( \left( \mu +\beta \right) \varphi \right) +a_{1}\rho
^{\alpha }\sin \left( \left( \mu -\alpha \right) \varphi \right) +a_{1}b\rho
^{\alpha +\beta }\sin \left( \left( \mu +\beta -\alpha \right) \varphi
\right)  \notag \\
& +a_{2}\rho ^{\beta }\sin \left( \left( \mu -\beta \right) \varphi \right)
+a_{3}b\rho ^{3\beta }\sin \left( \left( \mu -\beta \right) \varphi \right)
+a_{2}\rho ^{2\beta }\sin \left( \mu \varphi \right) \left( b-\frac{a_{3}}{%
a_{2}}\frac{\left\vert \sin \left( \left( \mu -2\beta \right) \varphi
\right) \right\vert }{\sin \left( \mu \varphi \right) }\right) .
\label{f-model-5}
\end{align}%
The thermodynamical restriction (\ref{Model 5 - tdr}) yields $0<2\beta -\mu
<\mu <1,$ so that by setting $\zeta =2\beta -\mu $ and $\xi =\mu $ in
function $g$ given by (\ref{g}), using (\ref{g-manje-g}) one has 
\begin{equation*}
\frac{\sin \left( \left( 2\beta -\mu \right) \varphi \right) }{\sin \left(
\mu \varphi \right) }<\frac{\left\vert \sin \frac{\left( \mu -2\beta \right)
\pi }{2}\right\vert }{\sin \frac{\mu \pi }{2}}.
\end{equation*}%
Therefore, again by (\ref{Model 5 - tdr}), one has that $b-\frac{a_{3}}{a_{2}%
}\frac{\left\vert \sin \left( \left( \mu -2\beta \right) \varphi \right)
\right\vert }{\sin \left( \mu \varphi \right) }>0,$ which, along with the
positivity of all other terms in (\ref{f-model-5}), implies that $f_{\rho
}\left( \varphi \right) >0$ if $\varphi \in \left( 0,\frac{\pi }{2}\right) .$

\paragraph{Model VI}

is obtained for $\gamma =\alpha +\beta ,$ $\mu =\beta ,$ and $\eta =\alpha ,$
so that function $f_{\rho },$ given by (\ref{f-od-fi}), reads%
\begin{equation}
f_{\rho }\left( \varphi \right) =\sin \left( \beta \varphi \right) +b\rho
^{\alpha }\sin \left( \left( \alpha +\beta \right) \varphi \right)
+a_{1}\rho ^{\alpha }\sin \left( \left( \beta -\alpha \right) \varphi
\right) +a_{1}b\rho ^{2\alpha }\sin \left( \beta \varphi \right) +a_{2}\rho
^{\alpha +\beta }\sin \left( \alpha \varphi \right) \left( b-\frac{a_{3}}{%
a_{2}}\right) .  \label{f-model-6}
\end{equation}%
The thermodynamical restriction (\ref{Model 6 - tdr}) yields $b-\frac{a_{3}}{%
a_{2}}>0,$ which, along with the positivity of all other terms in (\ref%
{f-model-6}), implies that $f_{\rho }\left( \varphi \right) >0$ if $\varphi
\in \left( 0,\frac{\pi }{2}\right) .$

\paragraph{Model VII}

is obtained for $\gamma =2\beta $ and $\mu =\eta =\beta ,$ so that function $%
f_{\rho },$ given by (\ref{f-od-fi}), reads%
\begin{equation}
f_{\rho }\left( \varphi \right) =\sin \left( \beta \varphi \right) +b\rho
^{\beta }\sin \left( 2\beta \varphi \right) +a_{1}\rho ^{\alpha }\sin \left(
\left( \beta -\alpha \right) \varphi \right) +a_{1}b\rho ^{\alpha +\beta
}\sin \left( \left( 2\beta -\alpha \right) \varphi \right) +a_{2}\rho
^{2\beta }\sin \left( \beta \varphi \right) \left( b-\frac{a_{3}}{a_{2}}%
\right) ,  \label{f-model-7}
\end{equation}%
The thermodynamical restriction (\ref{Model 7 - tdr}) yields $b-\frac{a_{3}}{%
a_{2}}>0,$ which, along with the positivity of all other terms in (\ref%
{f-model-7}), implies that $f_{\rho }\left( \varphi \right) >0$ if $\varphi
\in \left( 0,\frac{\pi }{2}\right) .$

\paragraph{Model VIII}

is obtained for $\gamma =2\alpha ,$ $\beta =\mu =\eta =\alpha ,$ $%
a_{1}+a_{2}=\bar{a}_{1},$ and $a_{3}=\bar{a}_{2},$ so that function $f_{\rho
},$ given by (\ref{f-od-fi}), reads%
\begin{equation}
f_{\rho }\left( \varphi \right) =\sin \left( \alpha \varphi \right) +b\rho
^{\alpha }\sin \left( 2\alpha \varphi \right) +\bar{a}_{1}\rho ^{2\alpha
}\sin \left( \alpha \varphi \right) \left( b-\frac{\bar{a}_{2}}{\bar{a}_{1}}%
\right) ,  \label{f-model-8}
\end{equation}%
The thermodynamical restriction (\ref{Model 8 - tdr}) yields $b-\frac{\bar{a}%
_{2}}{\bar{a}_{1}}>0,$ which, along with the positivity of all other terms
in (\ref{f-model-8}), implies that $f_{\rho }\left( \varphi \right) >0$ if $%
\varphi \in \left( 0,\frac{\pi }{2}\right) .$

\section{Calculation of the solution kernel \label{K-calc}}

In order to obtain the solution kernels, given by (\ref{fund sol 1}) and (%
\ref{fund sol 2}), the inverse Laplace transform (\ref{LIF}) will be
calculated using the Cauchy integral formula%
\begin{equation}
\oint_{\Gamma }\tilde{K}(x,s)\mathrm{e}^{st}\mathrm{d}s=0,\;\;x\in \mathbb{R}%
,\;t>0,  \label{Cauchy int th}
\end{equation}%
where $\Gamma $ is a closed curve containing the Bromwich path $\Gamma _{0}$
from the Laplace inversion formula (\ref{LIF}) and chosen differently
depending on the number and position of the branching points of function $%
\tilde{K},$ given by (\ref{k-lt}).

Branching points of function $\tilde{K}$ are points in which the function
under the square root is zero, i.e., in (\ref{k-lt}) either $\Phi _{\sigma
}(s)=0$ or $\Phi _{\varepsilon }(s)=0,$ $s\in 
\mathbb{C}
,$ with $\Phi _{\sigma }$ and $\Phi _{\varepsilon }$ given by (\ref%
{Burgers1-fiovi}) in the case of the first or by (\ref{Burgers2-fiovi}) in
the case of the second model class. Function $\Phi _{\varepsilon },$ except
for $s=0,$ does not have other zeros in the principal Riemann plane $\arg
s\in \left( -\pi ,\pi \right) ,$ since 
\begin{equation*}
\sum_{i=1}^{N}a_{i}s^{\alpha _{i}}\neq 0,\;\;s\in 
\mathbb{C}
,\;a_{i}\geq 0,\;\alpha _{i}\in \left[ 0,1\right) ,
\end{equation*}%
as proved in \cite{KOZ19}. Zeros of function 
\begin{equation*}
\Phi _{\sigma }(s)=1+a_{1}s^{\alpha }+a_{2}\,s^{\beta }+a_{3}\,s^{\gamma
},\;\;s\in 
\mathbb{C}
,
\end{equation*}%
with $a_{1},a_{2},a_{3}>0,$ $\alpha ,\beta \in \left( 0,1\right) ,$ $\gamma
\in \left( 0,2\right) ,$ and $\alpha <\beta <\gamma ,$ are analyzed in \cite%
{OZ-2}, where it is found that if $\gamma \in \left( 0,1\right) ,$ then
function $\Phi _{\sigma }$ has no zeros in the complex plane, which is valid
for Model I, while if $\gamma \in \left( 1,2\right) ,$ then the number and
position of zeros of function $\Phi _{\sigma }$\ is as follows:%
\begin{equation*}
\begin{tabular}{ll}
if $\func{Re}\Phi _{\sigma }\left( \rho ^{\ast }\right) <0,$ & 
\begin{tabular}{l}
then $\Phi _{\sigma }$ has no zeros in the complex plane;%
\end{tabular}
\\ 
if $\func{Re}\Phi _{\sigma }\left( \rho ^{\ast }\right) =0,$ & 
\begin{tabular}{l}
then $\Phi _{\sigma }$ has one negative real zero $-\rho ^{\ast }$;%
\end{tabular}
\\ 
if $\func{Re}\Phi _{\sigma }\left( \rho ^{\ast }\right) >0,$ & 
\begin{tabular}{l}
then $\Phi _{\sigma }$ has a pair of complex conjugated \\ 
zeros $s_{0}$ and $\bar{s}_{0}$ having negative real part;%
\end{tabular}%
\end{tabular}%
\end{equation*}%
where%
\begin{equation*}
\func{Re}\Phi _{\sigma }\left( \rho ^{\ast }\right) =1+a_{1}\left( \rho
^{\ast }\right) ^{\alpha }\cos \left( \alpha \pi \right) +a_{2}\left( \rho
^{\ast }\right) ^{\beta }\cos \left( \beta \pi \right) +a_{3}\left( \rho
^{\ast }\right) ^{\gamma }\cos \left( \gamma \pi \right) ,
\end{equation*}%
with $\rho ^{\ast }$ determined from $\func{Im}\Phi _{\sigma }\left( \rho
^{\ast }\right) =0,$ i.e., 
\begin{equation}
\frac{a_{1}\sin \left( \alpha \pi \right) }{a_{3}\left\vert \sin \left(
\gamma \pi \right) \right\vert }+\frac{a_{2}\sin \left( \beta \pi \right) }{%
a_{3}\left\vert \sin \left( \gamma \pi \right) \right\vert }\left( \rho
^{\ast }\right) ^{\beta -\alpha }=\left( \rho ^{\ast }\right) ^{\gamma
-\alpha },  \label{ro-zvezda-1}
\end{equation}%
which is valid for Models II - VII. In the case of Model VIII, zeros of
function 
\begin{equation*}
\Phi _{\sigma }\left( s\right) =1+\bar{a}_{1}s^{\alpha }+\bar{a}%
_{2}\,s^{2\alpha },\;\;s\in 
\mathbb{C}
,
\end{equation*}%
are as follows:%
\begin{equation*}
\begin{tabular}{ll}
\begin{tabular}{l}
if $\left( \frac{\bar{a}_{1}}{2\bar{a}_{2}}\right) ^{2}\geq \frac{1}{\bar{a}%
_{2}},$ or \\ 
if $\left( \frac{\bar{a}_{1}}{2\bar{a}_{2}}\right) ^{2}<\frac{1}{\bar{a}_{2}}
$ and $\frac{\bar{a}_{1}}{2\bar{a}_{2}}<\frac{\left\vert \cos \left( \alpha
\pi \right) \right\vert }{\sin \left( \alpha \pi \right) }\sqrt{\frac{1}{%
\bar{a}_{2}}-\left( \frac{\bar{a}_{1}}{2\bar{a}_{2}}\right) ^{2}},$%
\end{tabular}
& 
\begin{tabular}{l}
then $\Phi _{\sigma }$ has no zeros in the complex plane;%
\end{tabular}
\\ 
\begin{tabular}{l}
if $\left( \frac{\bar{a}_{1}}{2\bar{a}_{2}}\right) ^{2}<\frac{1}{\bar{a}_{2}}
$ and $\frac{\bar{a}_{1}}{2\bar{a}_{2}}=\frac{\left\vert \cos \left( \alpha
\pi \right) \right\vert }{\sin \left( \alpha \pi \right) }\sqrt{\frac{1}{%
\bar{a}_{2}}-\left( \frac{\bar{a}_{1}}{2\bar{a}_{2}}\right) ^{2}},$%
\end{tabular}
& 
\begin{tabular}{l}
then $\Phi _{\sigma }$ has one negative real zero $-\rho ^{\ast }$;%
\end{tabular}
\\ 
\begin{tabular}{l}
if $\left( \frac{\bar{a}_{1}}{2\bar{a}_{2}}\right) ^{2}<\frac{1}{\bar{a}_{2}}
$ and $\frac{\bar{a}_{1}}{2\bar{a}_{2}}>\frac{\left\vert \cos \left( \alpha
\pi \right) \right\vert }{\sin \left( \alpha \pi \right) }\sqrt{\frac{1}{%
\bar{a}_{2}}-\left( \frac{\bar{a}_{1}}{2\bar{a}_{2}}\right) ^{2}},$%
\end{tabular}
& 
\begin{tabular}{l}
then $\Phi _{\sigma }$ has a pair of complex conjugated \\ 
zeros $s_{0}$ and $\bar{s}_{0}$ having negative real part,%
\end{tabular}%
\end{tabular}%
\end{equation*}%
with $\rho ^{\ast }$ determined by%
\begin{equation}
\rho ^{\ast }=\left( \frac{b}{\sin \left( \alpha \pi \right) }\right) ^{%
\frac{1}{\alpha }}.  \label{ro-zvezda-2}
\end{equation}

Note that the branching point $s=0$ is due to the differentiation of
fractional order and that function $\tilde{K}$ does not have any
singularities other than branching points, justifying the use of the Cauchy
integral formula.

\subsection{Case 1.}

\subsubsection*{Function $\tilde{K},$ except for $s=0,$ has no other
branching points}

If function $\tilde{K}$ (\ref{k-lt}), except for $s=0,$ has no other
branching points, then the contour $\Gamma $ appearing in the Cauchy
integral formula (\ref{Cauchy int th}) is chosen as in Figure \ref{nemaTG}
and parametrized as in Table \ref{nemaTG-param}.

\noindent 
\begin{minipage}{\columnwidth}
\begin{minipage}[c]{0.4\columnwidth}
\centering
\includegraphics[width=0.7\columnwidth]{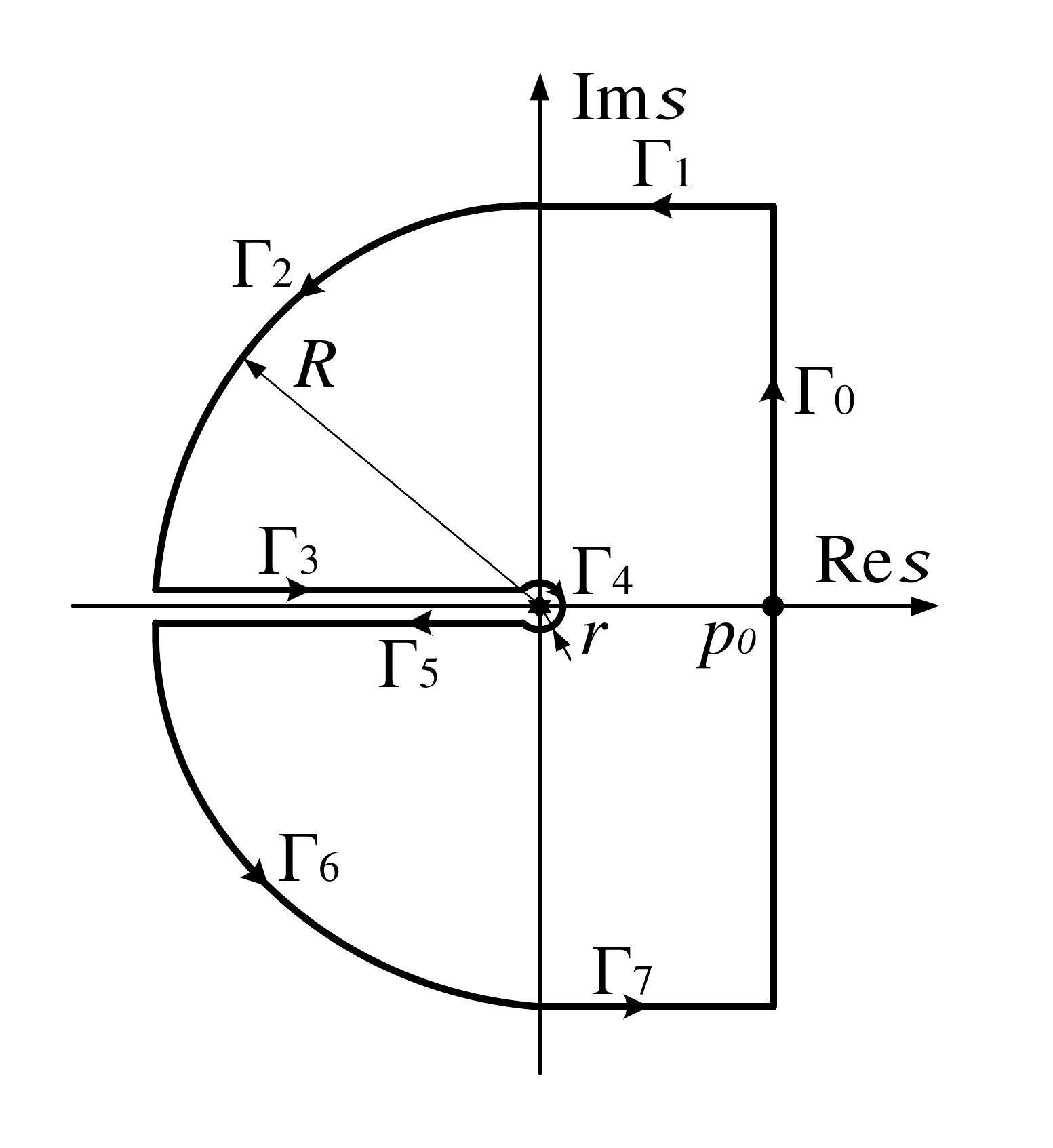}
\captionof{figure}{Integration contour $\Gamma$.}
\label{nemaTG}
\end{minipage}
\hfil
\begin{minipage}[c]{0.55\columnwidth}
\centering
\begin{tabular}{rll}
$\Gamma _{0}:$ & Bromwich path, &  \\ 
$\Gamma _{1}:$ & $s=p+\mathrm{i}R,$ & $p\in \left[ 0,p_{0}\right],\, p_0\geq 0$ arbitrary, \\ 
$\Gamma _{2}:$ & $s=R\mathrm{e}^{\mathrm{i}\varphi },$ & $\varphi \in \left[ 
\frac{\pi }{2},\pi \right] ,$ \\ 
$\Gamma _{3}:$ & $s=\rho \mathrm{e}^{\mathrm{i}\pi },$ & $\rho \in \left[ r,R%
\right] ,$ \\ 
$\Gamma _{4}:$ & $s=r\mathrm{e}^{\mathrm{i}\varphi },$ & $\varphi \in \left[ -\pi
,\pi \right] ,$ \\ 
$\Gamma _{5}:$ & $s=\rho \mathrm{e}^{-\mathrm{i}\pi },$ & $\rho \in \left[ r,R%
\right] ,$ \\
$\Gamma _{6}:$  & $s=R\mathrm{e}^{\mathrm{i}\varphi },$ & $\varphi \in \left[ 
-\pi, -\frac{\pi }{2} \right] ,$ \\
$\Gamma _{7}:$ & $s=p-\mathrm{i}R,$ & $p\in \left[ 0,p_{0}\right],\, p_0\geq 0$ arbitrary.  
\end{tabular}
\captionof{table}{Parametrization of integration contour $\Gamma$.}
\label{nemaTG-param}
\end{minipage}
\end{minipage}
\smallskip

The integrals along contours $\Gamma _{3},$ $\Gamma _{5},$ and $\Gamma _{0},$
calculated as%
\begin{eqnarray}
\lim_{\substack{ R\rightarrow \infty  \\ r\rightarrow 0}}\int_{\Gamma _{3}}%
\tilde{K}(x,s)\mathrm{e}^{st}\mathrm{d}s &=&\frac{1}{2}\int_{0}^{\infty }%
\sqrt{\frac{\Phi _{\sigma }(\rho \mathrm{e}^{\mathrm{i}\pi })}{\Phi
_{\varepsilon }(\rho \mathrm{e}^{\mathrm{i}\pi })}}\mathrm{e}^{\left\vert
x\right\vert \rho \sqrt{\frac{\Phi _{\sigma }(\rho \mathrm{e}^{\mathrm{i}\pi
})}{\Phi _{\varepsilon }(\rho \mathrm{e}^{\mathrm{i}\pi })}}}\mathrm{e}%
^{-\rho t}\mathrm{d}\rho ,  \label{int-gama-3} \\
\lim_{\substack{ R\rightarrow \infty  \\ r\rightarrow 0}}\int_{\Gamma _{5}}%
\tilde{K}(x,s)\mathrm{e}^{st}\mathrm{d}s &=&-\frac{1}{2}\int_{0}^{\infty }%
\sqrt{\frac{\Phi _{\sigma }(\rho \mathrm{e}^{-\mathrm{i}\pi })}{\Phi
_{\varepsilon }(\rho \mathrm{e}^{-\mathrm{i}\pi })}}\mathrm{e}^{\left\vert
x\right\vert \rho \sqrt{\frac{\Phi _{\sigma }(\rho \mathrm{e}^{-\mathrm{i}%
\pi })}{\Phi _{\varepsilon }(\rho \mathrm{e}^{-\mathrm{i}\pi })}}}\mathrm{e}%
^{-\rho t}\mathrm{d}\rho ,  \label{int-gama-5} \\
\lim_{\substack{ R\rightarrow \infty  \\ r\rightarrow 0}}\int_{\Gamma _{0}}%
\tilde{K}(x,s)\mathrm{e}^{st}\mathrm{d}s &=&2\pi \mathrm{i}K(x,t),
\label{int-gama-0}
\end{eqnarray}%
yield the solution kernel $K$ in the form (\ref{fund sol 1}) when used in
the Cauchy integral formula (\ref{Cauchy int th}), since the integrals along
all other contours will prove to be zero.

The following estimates will be used. According to (\ref{Burgers1-fiovi}),
respectively (\ref{Burgers2-fiovi}), after the substitution $s=\rho \mathrm{e%
}^{\mathrm{i}\varphi }$ is made, it is obtained that%
\begin{equation*}
\sqrt{\frac{\Phi _{\sigma }(s)}{\Phi _{\varepsilon }(s)}}\sim \left\{ 
\begin{tabular}{ll}
$\sqrt{\frac{a_{3}\,\rho ^{\gamma }\mathrm{e}^{\mathrm{i}\gamma \varphi }}{%
b\,\rho ^{\mu +\eta }\mathrm{e}^{\mathrm{i}\left( \mu +\eta \right) \varphi }%
}}=\sqrt{\frac{a_{3}}{b}}\rho ^{-\frac{\mu +\eta -\gamma }{2}}\mathrm{e}^{-%
\mathrm{i}\frac{\left( \mu +\eta -\gamma \right) \varphi }{2}},$ & for the
first model class,\medskip \\ 
$\sqrt{\frac{a_{3}\,\rho ^{\beta +\eta }\mathrm{e}^{\mathrm{i}\left( \beta
+\eta \right) \varphi }}{b\,\rho ^{\beta +\eta }\mathrm{e}^{\mathrm{i}\left(
\beta +\eta \right) \varphi }}}=\sqrt{\frac{a_{3}}{b}},$ & for the second
model class,%
\end{tabular}%
\right. \;\;\text{as}\;\;\rho \rightarrow \infty ,
\end{equation*}%
and therefore%
\begin{eqnarray}
\left\vert \sqrt{\frac{\Phi _{\sigma }(s)}{\Phi _{\varepsilon }(s)}}%
\right\vert &\sim &\left\{ 
\begin{tabular}{ll}
$\sqrt{\frac{a_{3}}{b}}\rho ^{-\frac{\mu +\eta -\gamma }{2}}\rightarrow 0,$
& for the first model class,\medskip \\ 
$\sqrt{\frac{a_{3}}{b}},$ & for the second model class,%
\end{tabular}%
\right. \;\;\text{as}\;\;\rho \rightarrow \infty ,  \label{moduo} \\
\arg \sqrt{\frac{\Phi _{\sigma }(s)}{\Phi _{\varepsilon }(s)}} &\sim
&\left\{ 
\begin{tabular}{ll}
$-\frac{\left( \mu +\eta -\gamma \right) \varphi }{2},$ & for the first
model class,\medskip \\ 
$0,$ & for the second model class,%
\end{tabular}%
\right. \;\;\text{as}\;\;\rho \rightarrow \infty .  \label{argument}
\end{eqnarray}

The integral along contour $\Gamma _{1}$ reads%
\begin{equation*}
\int_{\Gamma _{1}}\tilde{K}(x,s)\mathrm{e}^{st}\mathrm{d}s=\frac{1}{2}%
\int_{p_{0}}^{0}\sqrt{\frac{\Phi _{\sigma }(p+\mathrm{i}R)}{\Phi
_{\varepsilon }(p+\mathrm{i}R)}}\mathrm{e}^{-\left\vert x\right\vert \left(
p+\mathrm{i}R\right) \sqrt{\frac{\Phi _{\sigma }(p+\mathrm{i}R)}{\Phi
_{\varepsilon }(p+\mathrm{i}R)}}}\mathrm{e}^{\left( p+\mathrm{i}R\right) t}%
\mathrm{d}p,
\end{equation*}%
and since $p+\mathrm{i}R\sim R\,\mathrm{e}^{\mathrm{i}\frac{\pi }{2}},$ as $%
R\rightarrow \infty ,$ one has 
\begin{equation}
\lim_{R\rightarrow \infty }\left\vert \int_{\Gamma _{1}}\tilde{K}(x,s)%
\mathrm{e}^{st}\mathrm{d}s\right\vert \leq \frac{1}{2}\lim_{R\rightarrow
\infty }\int_{0}^{p_{0}}\left\vert \sqrt{\frac{\Phi _{\sigma }\left( R\,%
\mathrm{e}^{\mathrm{i}\frac{\pi }{2}}\right) }{\Phi _{\varepsilon }\left( R\,%
\mathrm{e}^{\mathrm{i}\frac{\pi }{2}}\right) }}\right\vert \mathrm{e}%
^{-\left\vert x\right\vert R\left\vert \sqrt{\frac{\Phi _{\sigma }(R\,%
\mathrm{e}^{\mathrm{i}\frac{\pi }{2}})}{\Phi _{\varepsilon }(R\,\mathrm{e}^{%
\mathrm{i}\frac{\pi }{2}})}}\right\vert \cos \left( \frac{\pi }{2}+\arg 
\sqrt{\frac{\Phi _{\sigma }(R\,\mathrm{e}^{\mathrm{i}\frac{\pi }{2}})}{\Phi
_{\varepsilon }(R\,\mathrm{e}^{\mathrm{i}\frac{\pi }{2}})}}\right) }\mathrm{e%
}^{pt}\mathrm{d}p.  \label{int-gama-1}
\end{equation}%
The use of (\ref{moduo}) and (\ref{argument}) in (\ref{int-gama-1}), due to $%
0<\frac{\mu +\eta -\gamma }{2}<1,$ yields%
\begin{equation*}
\lim_{R\rightarrow \infty }\left\vert \int_{\Gamma _{1}}\tilde{K}(x,s)%
\mathrm{e}^{st}\mathrm{d}s\right\vert \leq \frac{1}{2}\sqrt{\frac{a_{3}}{b}}%
\lim\limits_{R\rightarrow \infty }\int_{0}^{p_{0}}R^{-\frac{\mu +\eta
-\gamma }{2}}\mathrm{e}^{-\left\vert x\right\vert \sqrt{\frac{a_{3}}{b}}R^{1-%
\frac{\mu +\eta -\gamma }{2}}\cos \left( \left( 1-\frac{\mu +\eta -\gamma }{2%
}\right) \frac{\pi }{2}\right) }\mathrm{e}^{pt}\mathrm{d}p=0,
\end{equation*}%
for the first model class and choosing $p_{0}=0$%
\begin{equation*}
\lim_{R\rightarrow \infty }\left\vert \int_{\Gamma _{1}}\tilde{K}(x,s)%
\mathrm{e}^{st}\mathrm{d}s\right\vert \leq \frac{1}{2}\sqrt{\frac{a_{3}}{b}}%
\lim\limits_{R\rightarrow \infty }\int_{0}^{p_{0}}\mathrm{e}^{pt}\mathrm{d}%
p=0,
\end{equation*}%
for the second model class. Similar argumentation is valid for the integral
along $\Gamma _{7}$.

The integral along contour $\Gamma _{2}$ takes the form%
\begin{equation*}
\int_{\Gamma _{2}}\tilde{K}(x,s)\mathrm{e}^{st}\mathrm{d}s=\frac{1}{2}\int_{%
\frac{\pi }{2}}^{\pi }\sqrt{\frac{\Phi _{\sigma }(R\mathrm{e}^{\mathrm{i}%
\varphi })}{\Phi _{\varepsilon }(R\mathrm{e}^{\mathrm{i}\varphi })}}\mathrm{e%
}^{-\left\vert x\right\vert R\mathrm{e}^{\mathrm{i}\varphi }\sqrt{\frac{\Phi
_{\sigma }(R\mathrm{e}^{\mathrm{i}\varphi })}{\Phi _{\varepsilon }(R\mathrm{e%
}^{\mathrm{i}\varphi })}}}\mathrm{e}^{Rt\mathrm{e}^{\mathrm{i}\varphi }}%
\mathrm{i\,}R\mathrm{\,e}^{\mathrm{i}\varphi }\mathrm{d}\varphi ,
\end{equation*}%
so that%
\begin{equation}
\lim_{R\rightarrow \infty }\left\vert \int_{\Gamma _{2}}\tilde{K}(x,s)%
\mathrm{e}^{st}\mathrm{d}s\right\vert \leq \frac{1}{2}\lim_{R\rightarrow
\infty }\int_{\frac{\pi }{2}}^{\pi }R\left\vert \sqrt{\frac{\Phi _{\sigma }(R%
\mathrm{e}^{\mathrm{i}\varphi })}{\Phi _{\varepsilon }(R\mathrm{e}^{\mathrm{i%
}\varphi })}}\right\vert \mathrm{e}^{R\left( t\cos \varphi -\left\vert
x\right\vert \left\vert \sqrt{\frac{\Phi _{\sigma }(R\mathrm{e}^{\mathrm{i}%
\varphi })}{\Phi _{\varepsilon }(R\mathrm{e}^{\mathrm{i}\varphi })}}%
\right\vert \cos \left( \varphi +\arg \sqrt{\frac{\Phi _{\sigma }(R\mathrm{e}%
^{\mathrm{i}\varphi })}{\Phi _{\varepsilon }(R\mathrm{e}^{\mathrm{i}\varphi
})}}\right) \right) }\mathrm{d}\varphi .  \label{int-gama-2}
\end{equation}%
Using (\ref{moduo}) and (\ref{argument}) in (\ref{int-gama-2}), due to $0<%
\frac{\mu +\eta -\gamma }{2}<1$ and $\cos \varphi <0$ for $\varphi \in \left[
\frac{\pi }{2},\pi \right] ,$ yields%
\begin{equation*}
\lim_{R\rightarrow \infty }\left\vert \int_{\Gamma _{2}}\tilde{K}(x,s)%
\mathrm{e}^{st}\mathrm{d}s\right\vert \leq \frac{1}{2}\sqrt{\frac{a_{3}}{b}}%
\lim\limits_{R\rightarrow \infty }\int_{\frac{\pi }{2}}^{\pi }R^{1-\frac{\mu
+\eta -\gamma }{2}}\mathrm{e}^{R\left( t\cos \varphi -\left\vert
x\right\vert \sqrt{\frac{a_{3}}{b}}R^{-\frac{\mu +\eta -\gamma }{2}}\cos
\left( \left( 1-\frac{\mu +\eta -\gamma }{2}\right) \varphi \right) \right) }%
\mathrm{d}\varphi =0,
\end{equation*}%
for $\left( x,t\right) \in 
\mathbb{R}
\times \left[ 0,\infty \right) ,$ in the case of the first model class and 
\begin{equation*}
\lim_{R\rightarrow \infty }\left\vert \int_{\Gamma _{2}}\tilde{K}(x,s)%
\mathrm{e}^{st}\mathrm{d}s\right\vert \leq \frac{1}{2}\sqrt{\frac{a_{3}}{b}}%
\lim\limits_{R\rightarrow \infty }\int_{\frac{\pi }{2}}^{\pi }R\mathrm{\,e}%
^{R\left( t-\left\vert x\right\vert \sqrt{\frac{a_{3}}{b}}\right) \cos
\varphi }\mathrm{d}\varphi =0,\;\;\left\vert x\right\vert <\sqrt{\frac{b}{%
a_{3}}}t,
\end{equation*}%
for the second model class. Similar argumentation is valid for the integral
along $\Gamma _{6}$.

The integral along contour $\Gamma _{4}$:%
\begin{equation*}
\int_{\Gamma _{4}}\tilde{K}(x,s)\mathrm{e}^{st}\mathrm{d}s=\frac{1}{2}%
\int_{\pi }^{-\pi }\sqrt{\frac{\Phi _{\sigma }(r\mathrm{e}^{\mathrm{i}%
\varphi })}{\Phi _{\varepsilon }(r\mathrm{e}^{\mathrm{i}\varphi })}}\mathrm{e%
}^{-\left\vert x\right\vert r\mathrm{e}^{\mathrm{i}\varphi }\sqrt{\frac{\Phi
_{\sigma }(r\mathrm{e}^{\mathrm{i}\varphi })}{\Phi _{\varepsilon }(r\mathrm{e%
}^{\mathrm{i}\varphi })}}}\mathrm{e}^{rt\mathrm{e}^{\mathrm{i}\varphi }}%
\mathrm{i\,}r\mathrm{\,e}^{\mathrm{i}\varphi }\mathrm{d}\varphi
\end{equation*}%
tends to zero when $r\rightarrow 0$, since 
\begin{eqnarray*}
&&\lim_{r\rightarrow 0}\left\vert \int_{\Gamma _{4}}\tilde{K}(x,s)\mathrm{e}%
^{st}\mathrm{d}s\right\vert \leq \frac{1}{2}\lim_{r\rightarrow 0}\int_{-\pi
}^{\pi }r\left\vert \sqrt{\frac{\Phi _{\sigma }(r\mathrm{e}^{\mathrm{i}%
\varphi })}{\Phi _{\varepsilon }(r\mathrm{e}^{\mathrm{i}\varphi })}}%
\right\vert \mathrm{e}^{-\left\vert x\right\vert r\left\vert \sqrt{\frac{%
\Phi _{\sigma }(r\mathrm{e}^{\mathrm{i}\varphi })}{\Phi _{\varepsilon }(r%
\mathrm{e}^{\mathrm{i}\varphi })}}\right\vert \cos \left( \varphi +\arg 
\sqrt{\frac{\Phi _{\sigma }(r\mathrm{e}^{\mathrm{i}\varphi })}{\Phi
_{\varepsilon }(r\mathrm{e}^{\mathrm{i}\varphi })}}\right) }\mathrm{e}^{rt%
\mathrm{\cos }\varphi }\mathrm{d}\varphi \\
&&\qquad \leq \frac{1}{2}\left\{ 
\begin{tabular}{ll}
$\lim\limits_{r\rightarrow 0}\int_{-\pi }^{\pi }r^{1-\frac{\mu }{2}}\mathrm{e%
}^{-\left\vert x\right\vert r^{1-\frac{\mu }{2}}\cos \left( \left( 1-\frac{%
\mu }{2}\right) \varphi \right) }\mathrm{d}\varphi =0,$ & for the first
model class,\medskip \\ 
$\lim\limits_{r\rightarrow 0}\int_{-\pi }^{\pi }r^{1-\frac{\beta }{2}}%
\mathrm{e}^{-\left\vert x\right\vert r^{1-\frac{\beta }{2}}\cos \left(
\left( 1-\frac{\beta }{2}\right) \varphi \right) }\mathrm{d}\varphi =0,$ & 
for the second model class,%
\end{tabular}%
\right.
\end{eqnarray*}%
due to $\beta ,\mu <1$ and%
\begin{eqnarray*}
\left\vert \sqrt{\frac{\Phi _{\sigma }(s)}{\Phi _{\varepsilon }(s)}}%
\right\vert &\sim &\left\{ 
\begin{tabular}{ll}
$r^{-\frac{\mu }{2}},$ & for the first model class,\medskip \\ 
$r^{-\frac{\beta }{2}},$ & for the second model class,%
\end{tabular}%
\right. \;\;\text{as}\;\;r\rightarrow 0, \\
\arg \sqrt{\frac{\Phi _{\sigma }(s)}{\Phi _{\varepsilon }(s)}} &\sim
&\left\{ 
\begin{tabular}{ll}
$-\frac{\mu \varphi }{2},$ & for the first model class,\medskip \\ 
$-\frac{\beta \varphi }{2},$ & for the second model class,%
\end{tabular}%
\right. \;\;\text{as}\;\;\rho \rightarrow \infty .
\end{eqnarray*}

\subsubsection*{Function $\tilde{K},$ except for $s=0,$ has a negative real
branching point}

If function $\tilde{K}$ (\ref{k-lt}), except for $s=0,$ has a negative real
branching point $-\rho ^{\ast },$ determined by (\ref{ro-zvezda-1}) or (\ref%
{ro-zvezda-2}), then the contour $\Gamma $ appearing in the Cauchy integral
formula (\ref{Cauchy int th}) is chosen as in Figure \ref{negativnaTG} and
parametrized as in Table \ref{negativnaTG-param}.

\noindent 
\begin{minipage}{\columnwidth}
\begin{minipage}[c]{0.4\columnwidth}
\centering
\includegraphics[width=0.7\columnwidth]{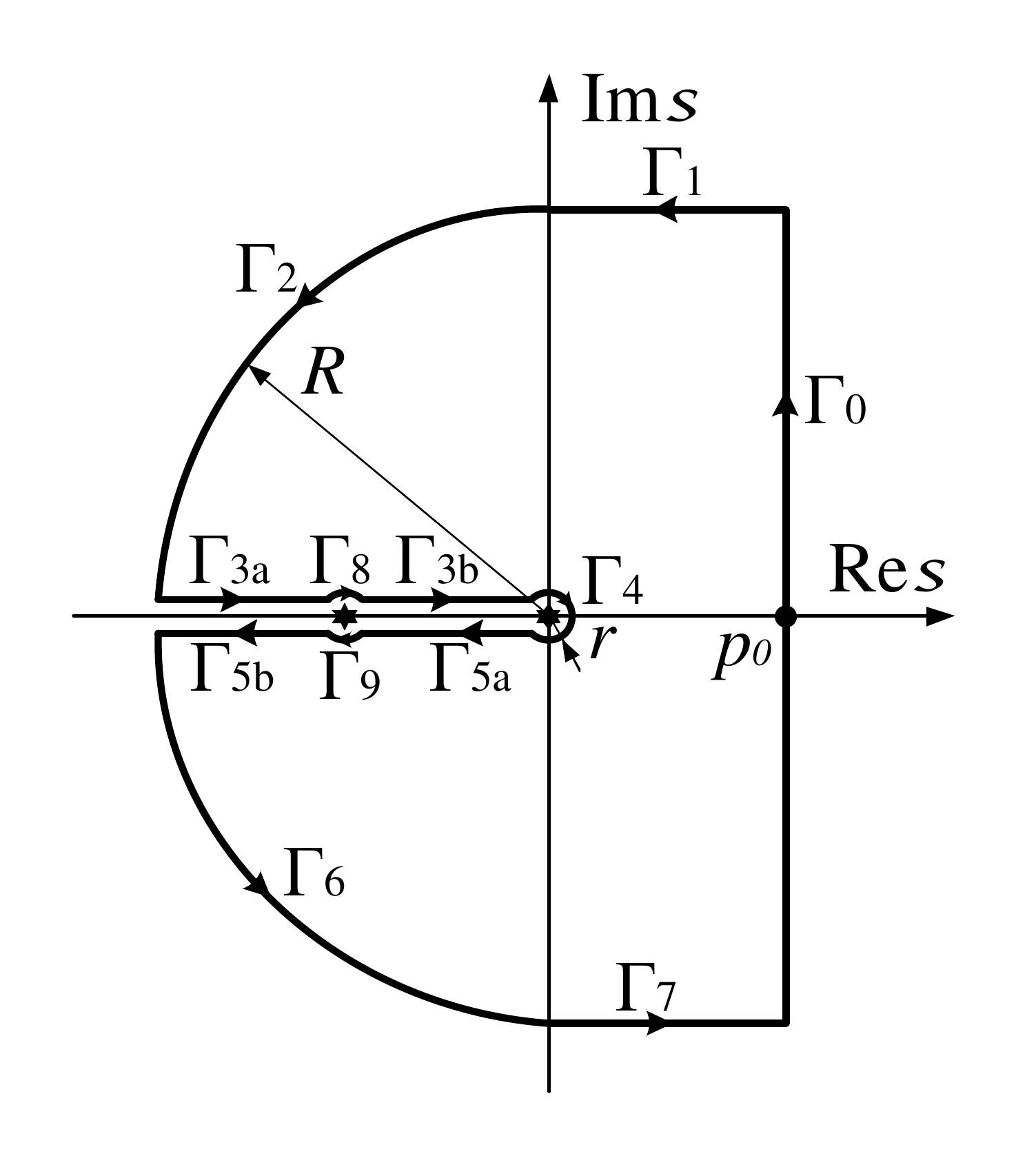}
\captionof{figure}{Integration contour $\Gamma$.}
\label{negativnaTG}
\end{minipage}
\hfil
\begin{minipage}[c]{0.55\columnwidth}
\centering
\begin{tabular}{rll}
$\Gamma _{0}:$ & Bromwich path, &  \\ 
$\Gamma _{1}:$ & $s=p+\mathrm{i}R,$ & $p\in \left[ 0,p_{0}\right], \, p_0\geq 0$ arbitrary, \\ 
$\Gamma _{2}:$ & $s=R\mathrm{e}^{\mathrm{i}\varphi },$ & $\varphi \in \left[ 
\frac{\pi }{2},\pi \right] ,$ \\ 
$\Gamma _{3a}:$ & $s=\rho \mathrm{e}^{\mathrm{i}\pi },$ & $\rho \in \left[ \rho^{*}+r,R\right] ,$ \\ 
$\Gamma _{3b}:$ & $s=\rho \mathrm{e}^{\mathrm{i}\pi },$ & $\rho \in \left[r, \rho^{*}-r\right] ,$ \\ 
$\Gamma _{4}:$ & $s=r\mathrm{e}^{\mathrm{i}\varphi },$ & $\varphi \in \left[ -\pi
,\pi \right] ,$ \\ 
$\Gamma _{5a}:$ & $s=\rho \mathrm{e}^{-\mathrm{i}\pi },$ & $\rho \in \left[r, \rho^{*}-r\right] ,$ \\
$\Gamma _{5b}:$ & $s=\rho \mathrm{e}^{-\mathrm{i}\pi },$ & $\rho \in \left[\rho^{*}+r,R\right] ,$ \\
$\Gamma _{6}:$  & $s=R\mathrm{e}^{\mathrm{i}\varphi },$ & $\varphi \in \left[ 
-\pi, -\frac{\pi }{2} \right] ,$ \\
$\Gamma _{7}:$ & $s=p-\mathrm{i}R,$ & $p\in \left[ 0,p_{0}\right],\, p_0\geq 0$ arbitrary, \\ 
$\Gamma _{8}:$ & $s-\rho^{*}\mathrm{e}^{\mathrm{i}\pi}=r\mathrm{e}^{\mathrm{i}\varphi },$ &  $\varphi \in \left[ 0,\pi \right],$ \\
$\Gamma _{9}:$ & $s-\rho^{*}\mathrm{e}^{-\mathrm{i}\pi}=r\mathrm{e}^{\mathrm{i}\varphi },$ &  $\varphi \in \left[-\pi,0 \right] .$
\end{tabular}
\captionof{table}{Parametrization of integration contour $\Gamma$.}
\label{negativnaTG-param}
\end{minipage}
\end{minipage}\smallskip

The integrals along contours $\Gamma _{3a}\cup \Gamma _{3b},$ $\Gamma
_{5a}\cup \Gamma _{5b},$ and $\Gamma _{0},$ when $r\rightarrow 0$ and $%
R\rightarrow \infty ,$ are the same integrals as (\ref{int-gama-3}), (\ref%
{int-gama-5}), and (\ref{int-gama-0}), thus yielding the solution kernel $K$
in the form (\ref{fund sol 1}) when used in the Cauchy integral formula (\ref%
{Cauchy int th}), since the integrals along contours $\Gamma _{1},$ $\Gamma
_{2},$ $\Gamma _{4},$ $\Gamma _{6},$ and $\Gamma _{7}$ already proved to be
zero, while the integrals along $\Gamma _{8}$ and $\Gamma _{9}$ will prove
to be zero.

Namely, the integral along $\Gamma _{8}$ reads%
\begin{equation*}
\int_{\Gamma _{8}}\tilde{K}(x,s)\mathrm{e}^{st}\mathrm{d}s=\frac{1}{2}%
\int_{\pi }^{0}\sqrt{\frac{\Phi _{\sigma }(\rho ^{\ast }\mathrm{e}^{i\pi }+r%
\mathrm{e}^{\mathrm{i}\varphi })}{\Phi _{\varepsilon }(\rho ^{\ast }\mathrm{e%
}^{i\pi }+r\mathrm{e}^{\mathrm{i}\varphi })}}\mathrm{e}^{-\left\vert
x\right\vert \left( \rho ^{\ast }\mathrm{e}^{i\pi }+r\mathrm{e}^{\mathrm{i}%
\varphi }\right) \sqrt{\frac{\Phi _{\sigma }(\rho ^{\ast }\mathrm{e}^{i\pi
}+r\mathrm{e}^{\mathrm{i}\varphi })}{\Phi _{\varepsilon }(\rho ^{\ast }%
\mathrm{e}^{i\pi }+r\mathrm{e}^{\mathrm{i}\varphi })}}}\mathrm{e}^{\left(
\rho ^{\ast }\mathrm{e}^{i\pi }+r\mathrm{e}^{\mathrm{i}\varphi }\right) t}%
\mathrm{i\,}r\mathrm{\,e}^{\mathrm{i}\varphi }\mathrm{d}\varphi ,
\end{equation*}%
so that 
\begin{equation*}
\lim_{r\rightarrow 0}\int_{\Gamma _{8}}\tilde{K}(x,s)\mathrm{e}^{st}\mathrm{d%
}s=\frac{1}{2}\mathrm{e}^{-\rho ^{\ast }t}\lim_{r\rightarrow 0}\int_{\pi
}^{0}\sqrt{\frac{\Phi _{\sigma }(\rho ^{\ast }\mathrm{e}^{i\pi }+r\mathrm{e}%
^{\mathrm{i}\varphi })}{\Phi _{\varepsilon }(\rho ^{\ast }\mathrm{e}^{i\pi
}+r\mathrm{e}^{\mathrm{i}\varphi })}}\mathrm{e}^{\left\vert x\right\vert
\rho ^{\ast }\sqrt{\frac{\Phi _{\sigma }(\rho ^{\ast }\mathrm{e}^{i\pi }+r%
\mathrm{e}^{\mathrm{i}\varphi })}{\Phi _{\varepsilon }(\rho ^{\ast }\mathrm{e%
}^{i\pi }+r\mathrm{e}^{\mathrm{i}\varphi })}}}\mathrm{i\,}r\mathrm{\,e}^{%
\mathrm{i}\varphi }\mathrm{d}\varphi =0,
\end{equation*}%
since 
\begin{equation*}
\lim_{r\rightarrow 0}\frac{\Phi _{\sigma }(\rho ^{\ast }\mathrm{e}^{i\pi }+r%
\mathrm{e}^{\mathrm{i}\varphi })}{\Phi _{\varepsilon }(\rho ^{\ast }\mathrm{e%
}^{i\pi }+r\mathrm{e}^{\mathrm{i}\varphi })}=\frac{\Phi _{\sigma }(\rho
^{\ast }\mathrm{e}^{i\pi })}{\Phi _{\varepsilon }(\rho ^{\ast }\mathrm{e}%
^{i\pi })}=0,
\end{equation*}%
because of $-\rho ^{\ast }$ being zero of function $\Phi _{\sigma }.$
Similar argumentation is valid for the integral along $\Gamma _{9}$.

\subsection{Case 2.}

\subsubsection*{Function $\tilde{K},$ except for $s=0,$ has a pair of
complex conjugated branching points}

If function $\tilde{K},$ except for $s=0,$ has a pair of complex conjugated
branching points with negative real part: $s_{0}=\rho _{0}\mathrm{e}^{%
\mathrm{i}\varphi _{0}}$ and $\bar{s}_{0}=\rho _{0}\mathrm{e}^{-\mathrm{i}%
\varphi _{0}}$, then the contour $\Gamma $ appearing in the Cauchy integral
formula (\ref{Cauchy int th}) is chosen as in Figure \ref{komplTG} and
parametrized as in Table \ref{komplTG-param}.

\noindent 
\begin{minipage}{\columnwidth}
\begin{minipage}[c]{0.4\columnwidth}
\centering
\includegraphics[width=0.7\columnwidth]{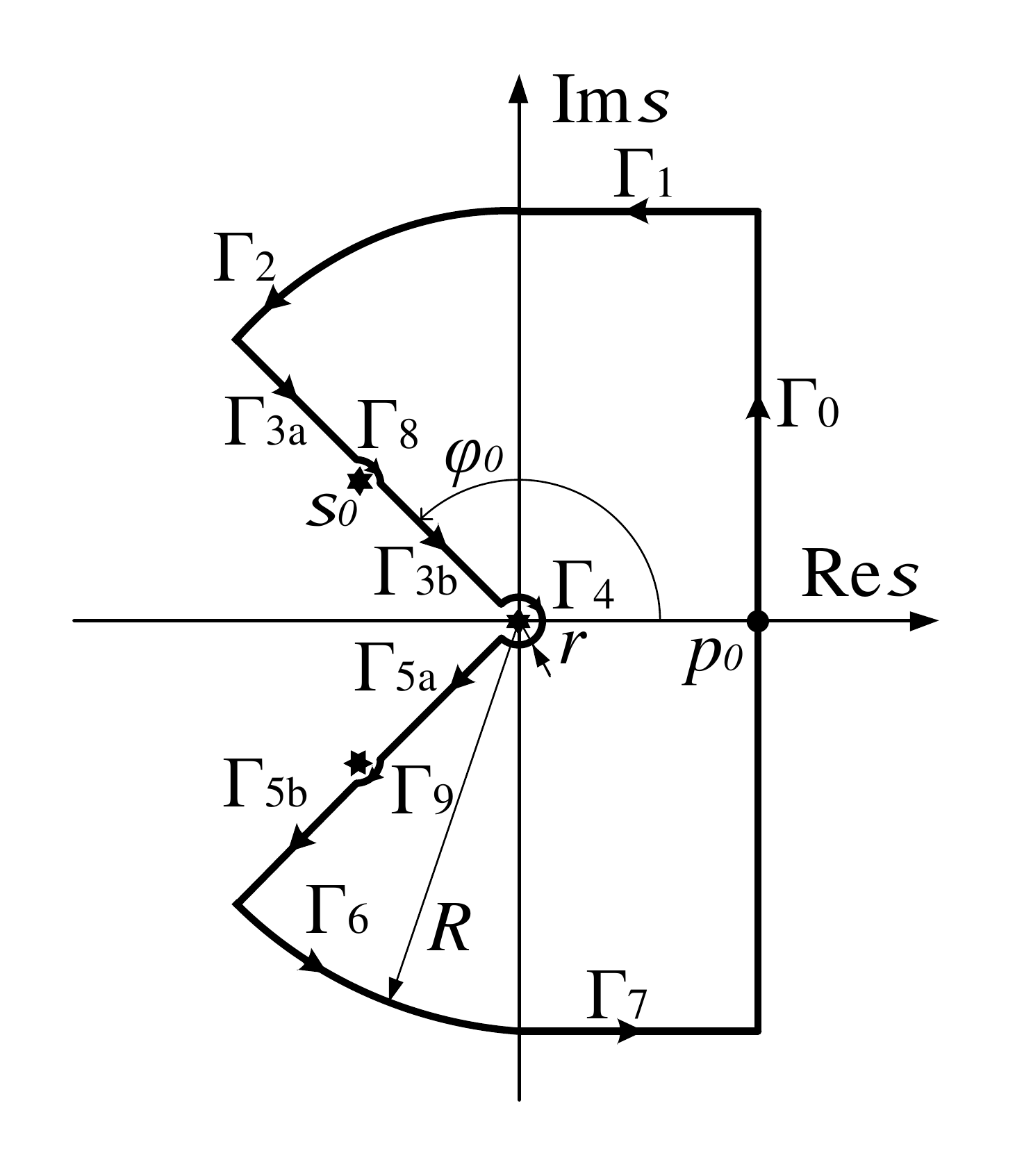}
\captionof{figure}{Integration contour $\Gamma$.}
\label{komplTG}
\end{minipage}
\hfil
\begin{minipage}[c]{0.55\columnwidth}
\centering
\begin{tabular}{rll}
$\Gamma _{0}:$ & Bromwich path, &  \\ 
$\Gamma _{1}:$ & $s=p+\mathrm{i}R,$ & $p\in \left[ 0,p_{0}\right],\, p_0\geq 0$ arbitrary, \\ 
$\Gamma _{2}:$ & $s=R\mathrm{e}^{\mathrm{i}\varphi },$ & $\varphi \in \left[ 
\frac{\pi }{2},\varphi_0 \right] ,$ \\ 
$\Gamma _{3a}:$ & $s=\rho \mathrm{e}^{\mathrm{i}\varphi_0 },$ & $\rho \in \left[ \rho_0+r,R\right] ,$ \\ 
$\Gamma _{3b}:$ & $s=\rho \mathrm{e}^{\mathrm{i}\varphi_0 },$ & $\rho \in \left[r, \rho_0-r\right] ,$ \\ 
$\Gamma _{4}:$ & $s=r\mathrm{e}^{\mathrm{i}\varphi },$ & $\varphi \in \left[ -\varphi_0,\varphi_0 \right] ,$ \\ 
$\Gamma _{5a}:$ & $s=\rho \mathrm{e}^{-\mathrm{i}\varphi_0 },$ & $\rho \in \left[r, \rho_0-r\right] ,$ \\
$\Gamma _{5b}:$ & $s=\rho \mathrm{e}^{-\mathrm{i}\varphi_0 },$ & $\rho \in \left[\rho_0+r,R\right] ,$ \\
$\Gamma _{6}:$  & $s=R\mathrm{e}^{\mathrm{i}\varphi },$ & $\varphi \in \left[ 
-\varphi_0, -\frac{\pi }{2} \right] ,$ \\
$\Gamma _{7}:$ & $s=p-\mathrm{i}R,$ & $p\in \left[ 0,p_{0}\right],\, p_0\geq 0$ arbitrary, \\
$\Gamma _{8}:$ & $s-s_0=r\mathrm{e}^{\mathrm{i}\varphi },$ & $\varphi \in \left[ -\varphi_0, \pi - \varphi_0 \right],$\\
$\Gamma _{9}:$ & $s-\bar{s}_0=r\mathrm{e}^{\mathrm{i}\varphi },$ & $\varphi \in \left[ -\pi + \varphi_0, \varphi_0 \right].$  
\end{tabular}
\captionof{table}{Parametrization of integration contour $\Gamma$.}
\label{komplTG-param}
\end{minipage}
\end{minipage}\smallskip

The solution kernel $K$ in the form (\ref{fund sol 2}) is obtained when the
integrals along contours $\Gamma _{3a}\cup \Gamma _{3b},$ $\Gamma _{5a}\cup
\Gamma _{5b},$ and $\Gamma _{0},$ calculated as 
\begin{eqnarray*}
\lim_{\substack{ R\rightarrow \infty  \\ r\rightarrow 0}}\int_{\Gamma
_{3a}\cup \Gamma _{3b}}\tilde{K}(x,s)\mathrm{e}^{st}\mathrm{d}s &=&-\frac{1}{%
2}\int_{0}^{\infty }\sqrt{\frac{\Phi _{\sigma }(\rho \mathrm{e}^{\mathrm{i}%
\varphi _{0}})}{\Phi _{\varepsilon }(\rho \mathrm{e}^{\mathrm{i}\varphi
_{0}})}}\mathrm{e}^{\mathrm{i}\varphi _{0}-\rho \mathrm{e}^{\mathrm{i}%
\varphi _{0}}\left( |x|\sqrt{\frac{\Phi _{\sigma }(\rho \mathrm{e}^{\mathrm{i%
}\varphi _{0}})}{\Phi _{\varepsilon }(\rho \mathrm{e}^{\mathrm{i}\varphi
_{0}})}}-t\right) }\mathrm{d}\rho , \\
\lim_{\substack{ R\rightarrow \infty  \\ r\rightarrow 0}}\int_{\Gamma
_{5a}\cup \Gamma _{5b}}\tilde{K}(x,s)\mathrm{e}^{st}\mathrm{d}s &=&\frac{1}{2%
}\int_{0}^{\infty }\sqrt{\frac{\Phi _{\sigma }(\rho \mathrm{e}^{-\mathrm{i}%
\varphi _{0}})}{\Phi _{\varepsilon }(\rho \mathrm{e}^{-\mathrm{i}\varphi
_{0}})}}\mathrm{e}^{-\mathrm{i}\varphi _{0}-\rho \mathrm{e}^{-\mathrm{i}%
\varphi _{0}}\left( |x|\sqrt{\frac{\Phi _{\sigma }(\rho \mathrm{e}^{-\mathrm{%
i}\varphi _{0}})}{\Phi _{\varepsilon }(\rho \mathrm{e}^{-\mathrm{i}\varphi
_{0}})}}-t\right) }\mathrm{d}\rho , \\
\lim_{\substack{ R\rightarrow \infty  \\ r\rightarrow 0}}\int_{\Gamma _{0}}%
\tilde{K}(x,s)\mathrm{e}^{st}\mathrm{d}s &=&2\pi \mathrm{i}K(x,t),
\end{eqnarray*}%
are used in the Cauchy integral formula (\ref{Cauchy int th}), since the
integrals along contours $\Gamma _{1},$ $\Gamma _{2},$ $\Gamma _{4},$ $%
\Gamma _{6},$ and $\Gamma _{7}$ already proved to be zero, while the
integrals along $\Gamma _{8}$ and $\Gamma _{9}$ will prove to be zero.

The integral along $\Gamma _{8}$ reads%
\begin{equation*}
\int_{\Gamma _{8}}\tilde{K}(x,s)\mathrm{e}^{st}\mathrm{d}s=\frac{1}{2}%
\int_{\varphi _{0}}^{-\pi +\varphi _{0}}\sqrt{\frac{\Phi _{\sigma }(s_{0}+r%
\mathrm{e}^{\mathrm{i}\varphi })}{\Phi _{\varepsilon }(s_{0}+r\mathrm{e}^{%
\mathrm{i}\varphi })}}\mathrm{e}^{-\left\vert x\right\vert \left( s_{0}+r%
\mathrm{e}^{\mathrm{i}\varphi }\right) \sqrt{\frac{\Phi _{\sigma }(s_{0}+r%
\mathrm{e}^{\mathrm{i}\varphi })}{\Phi _{\varepsilon }(s_{0}+r\mathrm{e}^{%
\mathrm{i}\varphi })}}}\mathrm{e}^{\left( s_{0}+r\mathrm{e}^{\mathrm{i}%
\varphi }\right) t}\mathrm{i\,}r\mathrm{\,e}^{\mathrm{i}\varphi }\mathrm{d}%
\varphi ,
\end{equation*}%
so that 
\begin{equation*}
\lim_{r\rightarrow 0}\int_{\Gamma _{8}}\tilde{K}(x,s)\mathrm{e}^{st}\mathrm{d%
}s=\frac{1}{2}\mathrm{e}^{s_{0}t}\lim_{r\rightarrow 0}\int_{\varphi
_{0}}^{-\pi +\varphi _{0}}\sqrt{\frac{\Phi _{\sigma }(s_{0}+r\mathrm{e}^{%
\mathrm{i}\varphi })}{\Phi _{\varepsilon }(s_{0}+r\mathrm{e}^{\mathrm{i}%
\varphi })}}\mathrm{e}^{\left\vert x\right\vert s_{0}\sqrt{\frac{\Phi
_{\sigma }(s_{0}+r\mathrm{e}^{\mathrm{i}\varphi })}{\Phi _{\varepsilon
}(s_{0}+r\mathrm{e}^{\mathrm{i}\varphi })}}}\mathrm{i\,}r\mathrm{\,e}^{%
\mathrm{i}\varphi }\mathrm{d}\varphi =0,
\end{equation*}%
since 
\begin{equation*}
\lim_{r\rightarrow 0}\frac{\Phi _{\sigma }(s_{0}+r\mathrm{e}^{\mathrm{i}%
\varphi })}{\Phi _{\varepsilon }(s_{0}+r\mathrm{e}^{\mathrm{i}\varphi })}=%
\frac{\Phi _{\sigma }(s_{0})}{\Phi _{\varepsilon }(s_{0})}=0,
\end{equation*}%
because of $s_{0}$ being zero of function $\Phi _{\sigma }.$ Similar
argumentation is valid for the integral along $\Gamma _{9}$.

\section*{Acknowledgment}

This work is supported by the Serbian Ministry of Education, Science and
Technological Development under grants $174005$ and $174024$, by the
Provincial Secretariat for Higher Education and Scientific Research under
grant $142-451-2384/2018$, as well as by FWO Odysseus project of Michael
Ruzhansky.


\end{document}